\documentclass[12pt]{iopart}

\begin{document}

\title{Notes on Dynamics of an External Cavity Semiconductor Lasers}
\author{S Behnia$^1$\footnote{Present address: Department of Physics,
Urmia University of Technology, Urmia, Iran}, Kh Mabhouti$^2$ and A
Jafari$^2$ A Akhshani$^{3,4}$}

\address{$^1$ Department of Physics, Urmia University of Technology, Urmia, Iran}
\address{$^2$ Department of Physics, Faculty of Science, Urmia University, Urmia, Iran}
\address{$^3$ Department of Physics, IAU, Orumieh Branch, Orumieh, Iran}
\address{$^4$ School of Physics, Universiti Sains Malaysia, 11800 USM, Penang, Malaysia}
\ead{$sohrab\_behnia@yahoo.com$}

\begin{abstract}
Dynamics of external cavity semiconductor lasers is known to be a
complex and uncontrollable phenomenon. Due to the lack of
experimental studies on the nature of the external cavity
semiconductor lasers, there is a need to theoretically clarify laser
dynamics. The stability of laser dynamics in the present paper, is
analyzed through plotting the Lyapunov exponent spectra, bifurcation
diagrams, phase portrait and electric field intensity time series.
The analysis is preformed with respect to applied feedback phase
$C_p$, feedback strength $\eta$ and the pump current of the laser.
The main argument of the paper is to show that the laser dynamics
can not be accounted for through simply a bifurcation diagram and
single-control parameter. The comparison of the obtained results
provides a very detailed picture of the qualitative changes in laser
dynamics.
\end{abstract}

\pacs{42.65.Sf, 42.55.Px, 42.55.Px, 05.45.Vx}

 %Keywords required only for MST, PB, PMB, PM, JOA, JOB?

\vspace{2pc} \noindent{\it Keywords}: Short external cavity
semiconductor Lasers, Optical feedback, Feedback strength, Pumping
current, Lyapunov exponent spectrum, Bifurcation.

 %Uncomment for Submitted to journal title message
\submitto{\JPA}
 %Comment out if separate title page not required
\maketitle

\section{Introduction}
Semiconductor lasers are ubiquitous in the modern world. They are
the principal source of coherent light in optical communications and
ultra-fast optical processing, they are used in optical storage
devices,in laser pointers, to cite only a few
applications~\cite{L2,SCO1,SCO2,SCO3}. Semiconductor lasers provide
one of the best physical systems for studying nonlinear dynamic
phenomena. Many researchers have investigated
the chaotic behavior of several laser systems~\cite{L,L0,L1}.\\
Semiconductor laser has many advantages, such as small size, easy
integration, compactness, low cost and convenience of operation.
Therefore, they are preferable to any other types of
lasers in the field of optical telecommunications.\\
External cavity semiconductor lasers (ECSLs) are an integral part of
high speed chaos based communication systems~\cite{CB,SC3}. Hence,
ECSLs have been a subject of extensive research~\cite{ex,SC4,SC5}.
Understanding the influence of delayed optical feedback on the
behavior of ECSLs is of great relevance for technological
applications.\\
Different studies have been conducted to characterize or manipulate
of ECSLs; long external cavity~\cite{SC5,lec} and short external
cavity~\cite{cte1,sec1,sec2}. Because of advantages in the short
cavity regime~\cite{cte1,Time}, in this paper, we just focus on the
short external cavity semiconductor lasers as a dynamical system.
The main model in the understanding of the dynamics of these lasers
is the well-known Lang-Kobayashi (LK) equations, consisting of two
delay differential equations (DDEs)~\cite{DDE1} for the complex
electrical field $E$ and the carrier number density
$N$~\cite{EC2,EC3}. These equations closely describe the chaotic
behavior of the physical system and confirm the presence of chaotic
regimes~\cite{EC1}. A large part of these studies have been made
possible using numerical continuation methods which allow one to
find and follow numerical solution to the rate equations irrespective of their stability.\\
Our main point here is that one is likely to miss important
phenomena if one just considers a bifurcation diagram. This is why
we show the behavior of the laser for a single value of the control
parameter in different ways: by Lyapunov exponents and time series.
This allows us to present a consistent overall picture of the
dynamics at the same time. In this study parameters such as feedback
phase, feedback strength, and pump current in the short external
cavity are used as control parameters to obtain different dynamical
regimes, including periodic and quasi periodic (QP)~\cite{QP},
regular pulse packages (RPP)~\cite{Time}, and chaotic (CH)
behavior~\cite{Chaos}.

\section{The Laser model}
In the early 1980s, Lang-Kobayashi (LK) proposed model semiconductor
lasers\\~\cite{LK1,LK2}. The $LK$ equations are model equations that
have been used extensively in the past to describe a semiconductor
laser subject to feedback from an external cavity~\cite{LK2} where
they used DDEs with the advantage that have an infinite-dimensional
phase space~\cite{DDE1}. For the (complex) electric field E and
inversion N, we write the $LK$ equations as the dimensionless and
compact set of equations~\cite{SC5}:
\begin{equation}
\frac{dE}{dt}=(1+i\alpha)NE+\eta E(t-\tau)e^{-iC_p}
\end{equation}
\begin{equation}
T\frac{dN}{dt}=P-N-(1-2N)\mid E\mid^2.
\end{equation}
The parameters in the above-mentioned equations describe the line
width enhancement factor $\alpha$, the feedback strength $\eta$ ,
the $2\pi$-periodic feedback phase $C_p$, the ratio between carrier,
photon lifetime $T$ and the pump current ${\bf P}$. Equations (1),
(2) describe a semiconductor laser with external optical feedback.
In these equations, the time is normalized to the cavity photon
lifetime $(~1 ps)$ and T is the ratio of the carrier lifetime $(1
ns)$ to the photon lifetime~\cite{Time}. In this study the
parameters are chosen not only to elucidate the dynamical structure
but also correspond fairly well to the experimental conditions. The
external round trip time $\tau$ is also normalized to the photon
lifetime. The remaining parameters, however, are held fixed at
$T=1710$, $\tau=70$, $\alpha=5.0$~\cite{cte1}. In this paper the
stability of an electric field intensity $\mid E\mid^2$ is studied
versus $C_p$, $\eta$ , and {\bf P}. Where all the parameters are
easily accessible in experiments~\cite{cte1,Exp1}.
\section{Stability analysis}
\subsection{Bifurcation diagrams}
 Bifurcation means a qualitative change in the dynamical behavior of
a system when a parameter of the system is varied. A bifurcation
diagram provides a useful insight into the transition between
different types of motion that can occur as one parameter of the
system alters~\cite{Bif1}. It enables one to study the behavior of
the system on a wide range of an interested control parameter. In
this paper the dynamical behavior of the system is studied through
plotting the bifurcation diagrams of the $\mid E \mid^2$ versus
$C_p$, $\eta$, and {\bf P} as control parameters. This procedure
continued by increasing the control parameters, and the new
resulting points were plotted in the bifurcation diagram versus the
new control parameter.
\subsection{Lyapunov exponent spectrum}
Lyapunov exponents and entropy measures, on the other hand can be
considered as `dynamic' measures of attractor complexity and they
are called `time average'~\cite{Bif1}. The Lyapunov exponent
$\lambda$ is useful for distinguishing various orbits. Lyapunov
Exponents quantify sensitivity of the system to initial conditions
and give a measure of predictability. The Lyapunov exponent is a
measure of the rate at which the trajectories separate one from
another. A negative exponent implies that the orbits approach to a
common fixed point. A zero exponent means that the orbits maintain
their relative positions; they are on a stable attractor. Finally, a
positive exponent implies that the orbits are on a chaotic
attractor, so the presence of a positive Lyapunov
exponent indicates chaos. The Lyapunov exponent is defined as follows:\\
Consider two nearest neighboring points in phase space at time $0$
and $t$, with distances of the points in the $ith$ direction
$\|{\delta}x_i(0)\|$, and $\|{\delta}x_i(t)\|$, respectively. The
Lyapunov exponent is then defined by the average growth rate
$\lambda_i$ of the initial distance,
\begin{equation}
{\lambda_i}=
\lim_{t{\rightarrow}{\infty}}{\frac{1}{t}}\ln{\frac{\|{\delta}x_i(t)\|}{\|{\delta}x_i(0)\|}}
\end{equation}
The existence of a positive Lyapunov exponent is the indicator of
chaos showing neighboring points with infinitesimal differences at
the initial state abruptly separate from each other in the $ith$
direction~\cite{Lya1}. Using the algorithm of Wolf~\cite{Wolf}, the
Lyapunov exponent was calculated versus a given control parameter.
Then the value of the control parameter increased a little and the
Lyapunov exponent was calculated for the new control parameter. By
continuing this procedure Lyapunov exponent spectrum of the system
was plotted versus the control parameter.
\section{Results and Discussions} The absence of employing direct
mathematical methods in study of ECSLs with respect to the
variations of the parameters $C_p$, $\eta$, and {\bf P} easily could
be  solved by considering the bifurcation diagram and Lyapunov
exponent spectrum. The output intensity $(\mid E \mid^2)$ dynamics
in the ECSLs, has been studied by considering both the variation of
each parameter ($\eta$, $C_p$,\textbf{P}) and the two other ($\eta$,
 $C_p$, \textbf{P}) as a initial condition of ECSLs. The process  is divided into three categories as defined below.\\
\textbf{\textit{A- The effect of the feedback strength variations:}}
The effect of the increasing of the $\eta$ in the laser dynamics
could be divided into the two sections of low regime $\eta$ (such as
$\eta <0.0605$) and high regime $\eta$ (such as $\eta
>0.0727$)~\cite{cte1}. In the low $\eta$ as it was expected $\mid E
\mid^2$ dynamics reveals chaotic (CH), periodic (P1,P2,...), and
quasi periodic (QP) behaviors.  By the same token in the high regime
$\mid E \mid^2$ dynamics reveals a regular plus package (RPP), QP
and P1 behaviors (see Fig. 1). The basic role of the $\eta$ in laser
dynamics is to displace of the periodic behavior windows, which can
be presented by the Lyapunov exponent spectrum and the bifurcation diagrams.\\

\textbf{ \textit{B- The effect of the feedback phase variations:}}
The optical feedback phase is of a particular cyclic nature.
Starting from a certain initial status, a variation of the $C_p$ by
$2\pi$ must turn back to its initial status~\cite{cte1}. To
illustrate the effect of  $C_p$ and its sensitive influence on the
output intensity, $\mid E\mid^2$, the study should focus on the two,
Low and high regimes. In the low regime of $\eta$, the output
intensity $\mid E \mid^2$ presents a P1 and  P2, the chaotic
behavior and QP (see Fig. 2). The appearance of the regular pulse
packages (RPP) and disappearing of both the chaotic behavior and
higher period (P2, P3,...) is the natural function of the output
intensity $\mid E \mid^2$ in the high regime of the $C_p$ (see Fig.
3). The previous studies predicted bifurcation up the mode 10 (such
as P1, P2, ...). However, based on the result of the present study,
bifurcation can not exceed 7 mode (see Fig. 4).\\
\textbf{\textit{C- The effect of  pumping current variations:}}
 The bifurcation diagram in variations of pump current follows the threshold
value with respect to the $\eta$ and $C_p$. Where by considering the
low $\eta$, as can be seen in Fig. 5a, the $\mid E\mid^2$ is
initially stable  with a P1 which undergoes early cascades of
period-doubling to chaos. Chaotic window in bifurcation diagram
varies based on the  $C_p$, as shown in Fig. 5(a-c).\\
In high $\eta$, as pictured in Fig. 6a, the $\mid E\mid^2$ is
initially( at first ) in P1 dynamics, which by increasing the
\textbf{P}, it undergoes RPP windows. As shown in Fig. 6(a-c),
threshold of RPP is directly depended to $C_p$ with a higher and
minor growth rate. At low $\eta$, variations of periodic window and
the increase of the chaotic window are the results of the increase
of the pump current depicted in Fig. 7(a-c). The obtained results
confirmed by plotting the time and phase portrait too (Fig. 8 and
Fig. 9). Fig. 8(a-b) and Fig. 9(a-b), verify the increase of the
chaotic window. Fig. 8(c-d) and Fig. 9(c-d), on the other hand
confirm variations of periodic window. Also based on the result the
increase of the RPP windows and the disappearance of the QP in high
$\eta$ is the result of the increasing of the pump current (see Fig.
10 and Fig. 11). The overall picture of $\mid E \mid^2$ dynamics has
been presented in Fig. 12.
\section{Summery and Outlook}
Previous studies demonstrated have not provided sufficient insists
towards the nature of ECSLs dynamics. Thus, there is a need to
theoretically clarify its dynamics~\cite{Time}. As noted before,
bifurcation diagram can not reveal the hidden aspects of Laser
dynamics~\cite{Hinted,SC5}. Therefore, employing the Lyapunov
exponent spectrum method can cover this methodological deficiency.
The obtained results shed some lights on control process of the
ECSLs dynamics. As a conclusions, it can be stated that $\eta$ shows
its influence on Laser dynamics by selecting a method of transfer
form stable status to unstable status~\cite{cte1} such as (Periodic,
QP and chaotic) behavior or (RPP, QP and P1). Therefor $\eta$  can
be regarded as the most important factor in stability of the Laser.
Where $C_p$ acts as a selection method in the scenario. Pump current
is the other important factor in Laser dynamics, an essential factor
which can influence the ECSLs dynamics from practical point of view.
The use of pump current for controlling the ECSLs dynamics can also
be considered a new field for further investigations.

\clearpage
\newpage
 {\bf Figure Captions}\\

Fig.1. Development of the dynamics for $\mid E \mid^2$(${\bf P}=0.8$
and $C_p=1$) (a)- Bifurcation diagram (b)-Lyapunov exponent
spectrum.

Fig.2.  Illustration of the effect of the $C_p$ variation on $\mid E
\mid^2$(${\bf P}=0.8$ and $\eta=0.036$) (a)- Bifurcation diagram
(b)-Lyapunov exponent spectrum.

Fig.3. Illustration of the effect of the $C_p$ variation on $\mid E
\mid^2$(${\bf P}=0.8$ and $\eta=0.090$) (a)- Bifurcation diagram
(b)-Lyapunov exponent spectrum.

Fig.4. A comparative description of various periodic modes(${\bf
P}=0.8$) (a)- $\eta=0.042$, (b)- $\eta=0.0455$, (c)- $\eta=0.048$.

Fig.5. The analysis of $\mid E \mid^2$ dynamics($\eta=0.0455$), (a)-
$C_p=-1$, (b)- $C_p=0$, (c)- $C_p=1$.

Fig.6. The analysis of $\mid E \mid^2$ dynamics($\eta=0.115$), (a)-
$C_p=-1$, (b)- $C_p=0$, (c)- $C_p=1$.

Fig.7. The role of pump currents as initial condition in periodic
and chaotic windows ($\eta=0.0455$) (a),(b) ${\bf P}=0.6$; (c), (d)
${\bf P}=0.7$; (e), (f)- ${\bf P}=1.4$.

Fig.8. Time series of $\mid E \mid^2$($\eta=0.0455$)(a)- $C_p=-1$
and ${\bf P}=0.6$ (QP), (b)- $C_p=-1$ and ${\bf P}=0.7$ (Chaotic),
(c)- $C_p=-2.835$ and ${\bf P}=0.7$ (P4), (d)- $C_p=-1$ and ${\bf
P}=1.4$ (P2).

Fig.9. Phase diagram of $\mid E \mid^2$ ($\eta=0.0455$) (a)-
$C_p=-1$ and ${\bf P}=0.6$ (QP), (b)- $C_p=-1$ and ${\bf P}=0.7$
(Chaotic), (c)- $C_p=-2.835$ and ${\bf P}=0.7$ (P4), (d)- $C_p=-1$
and ${\bf P}=1.4$ (P2).

Fig 10. The effect of ${\bf P}$ in increasing the RPP
windows($\eta=0.042$) (a)-${\bf P}=0.8$ (b)- ${\bf P}=1.4$.

Fig 11. Increasing RPP windows on $\mid E \mid^2$ ($C_p=0$,
$\eta=0.09$ and ${\bf P}=0.8$) (a)- Time series (b)- Phase diagram.

Fig 12. The overall picture of $\mid E\mid^2$ dynamics.

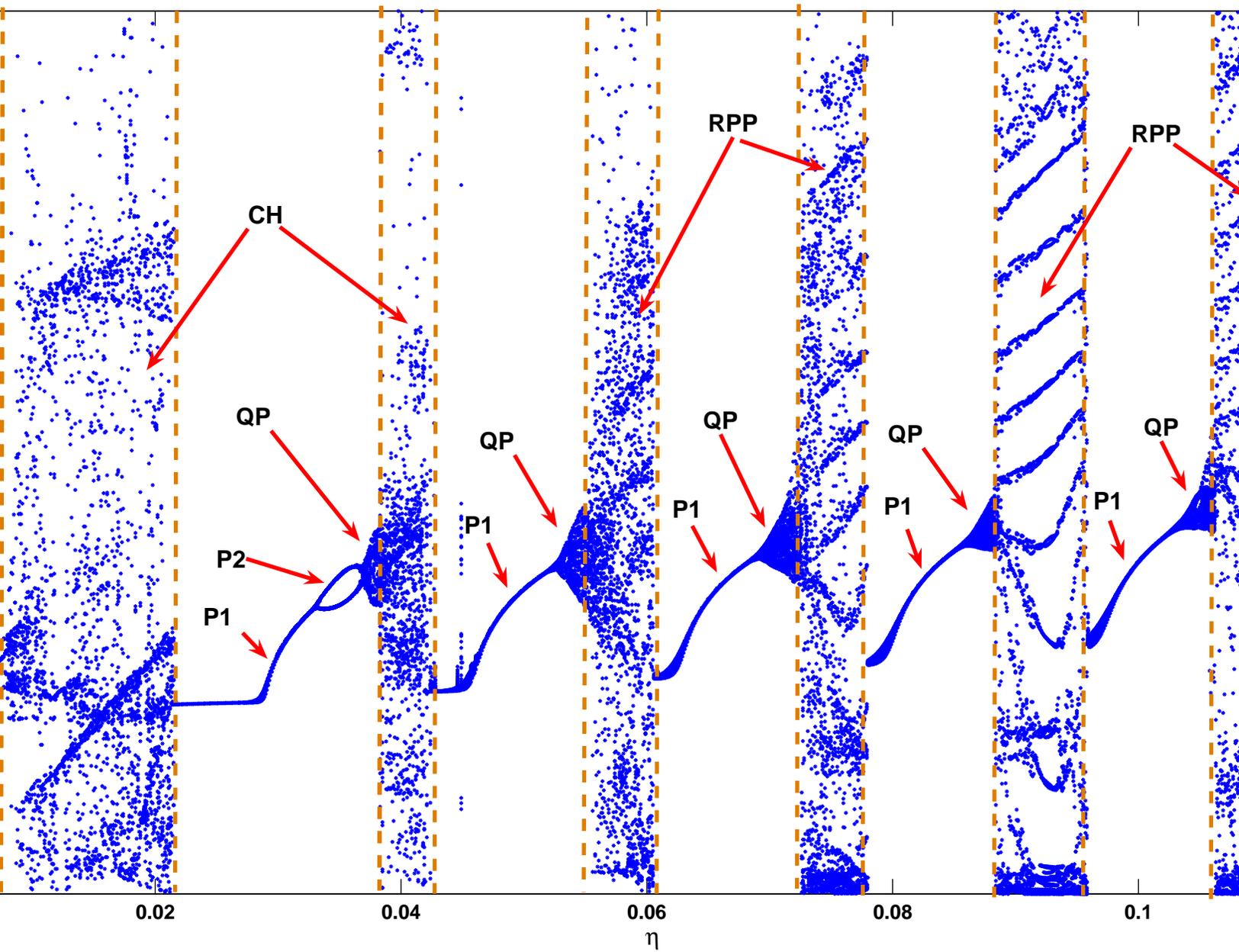
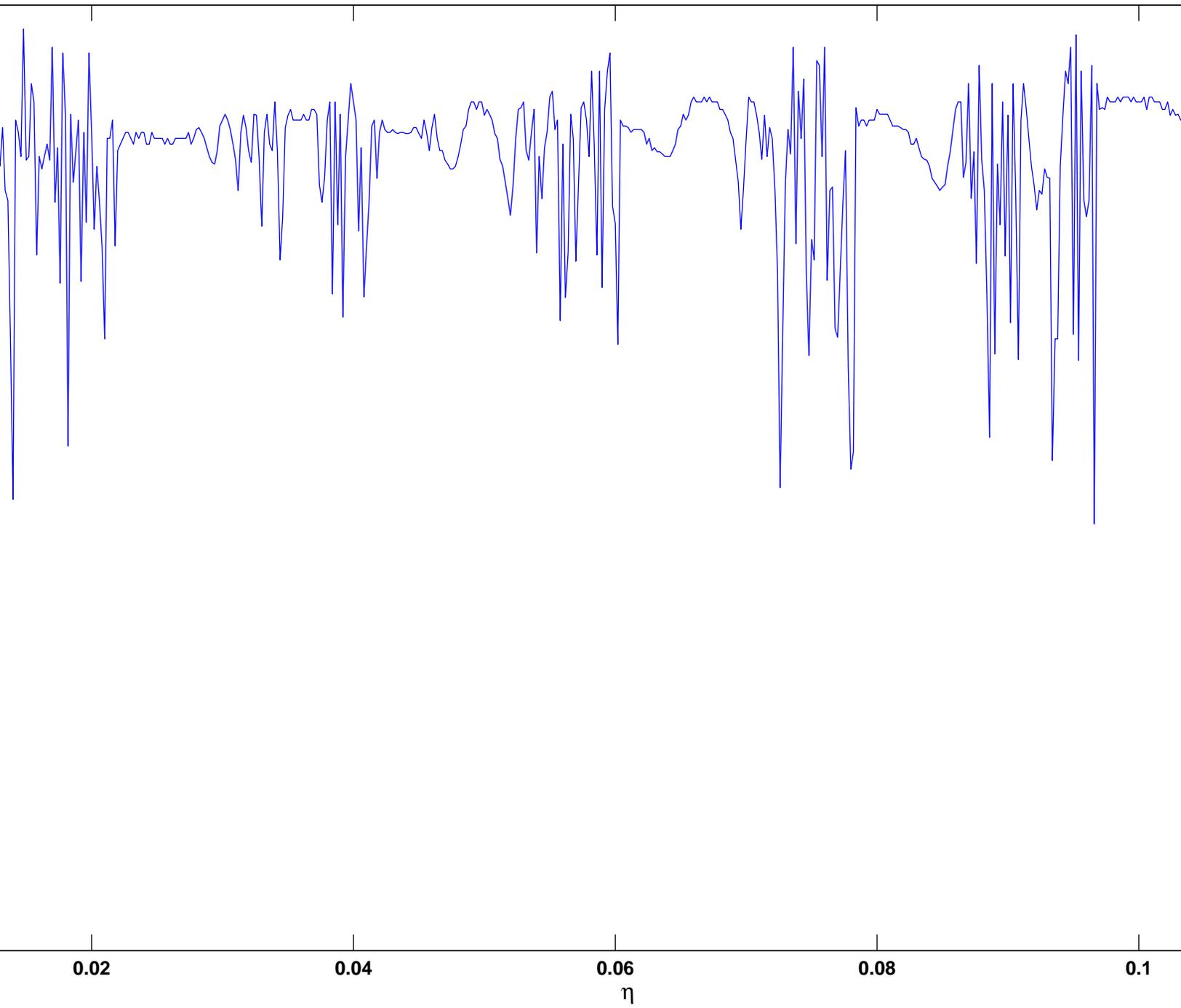
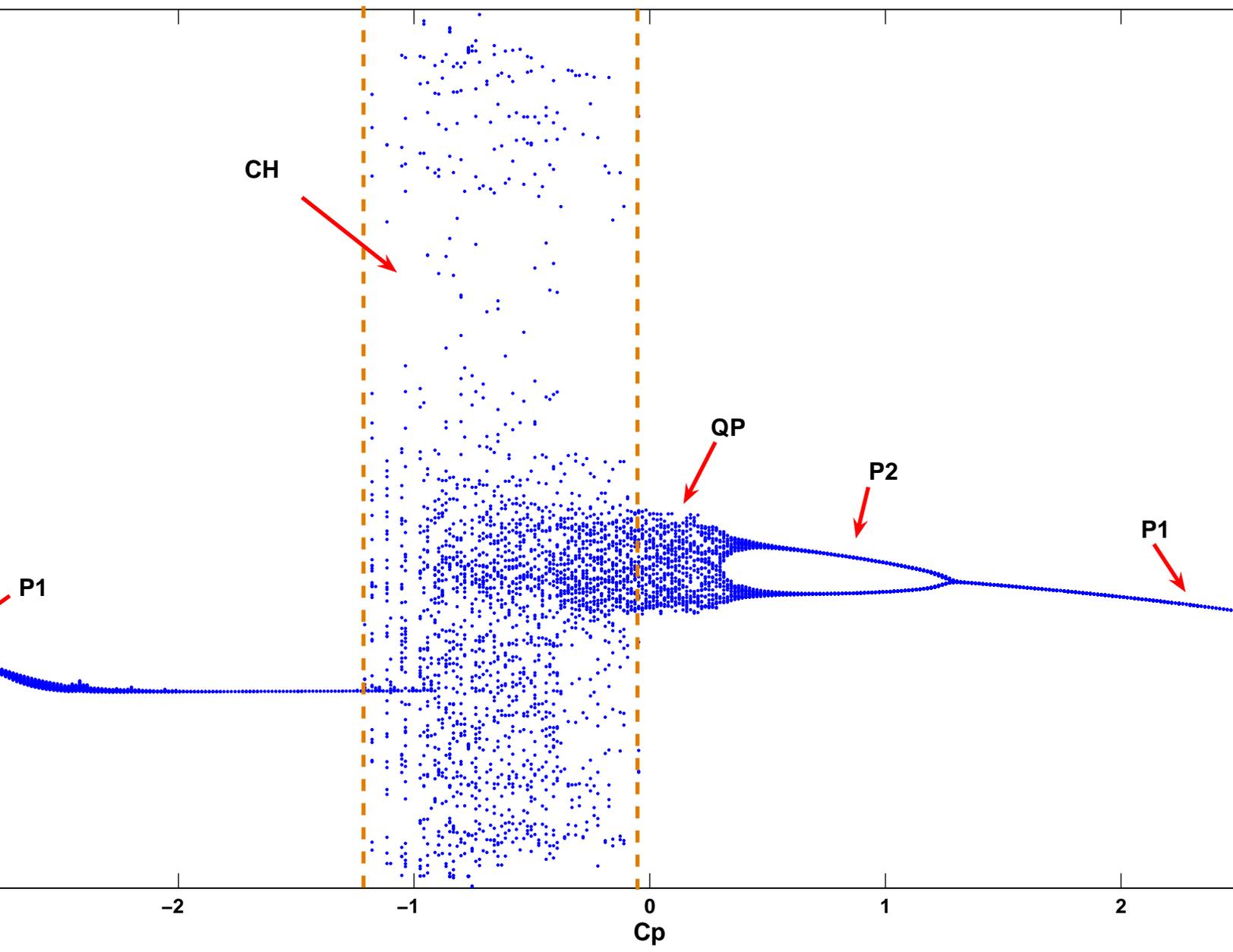
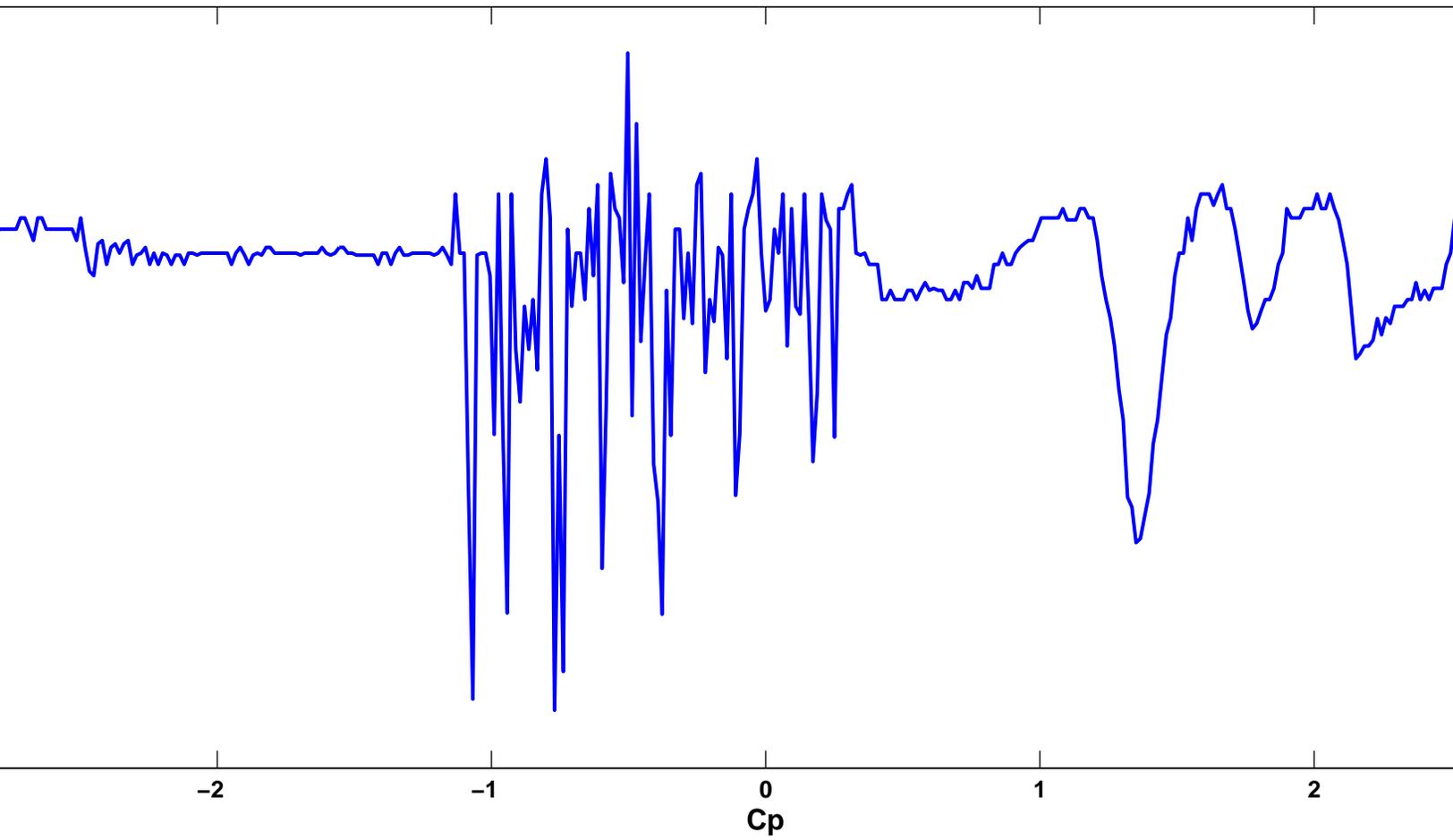
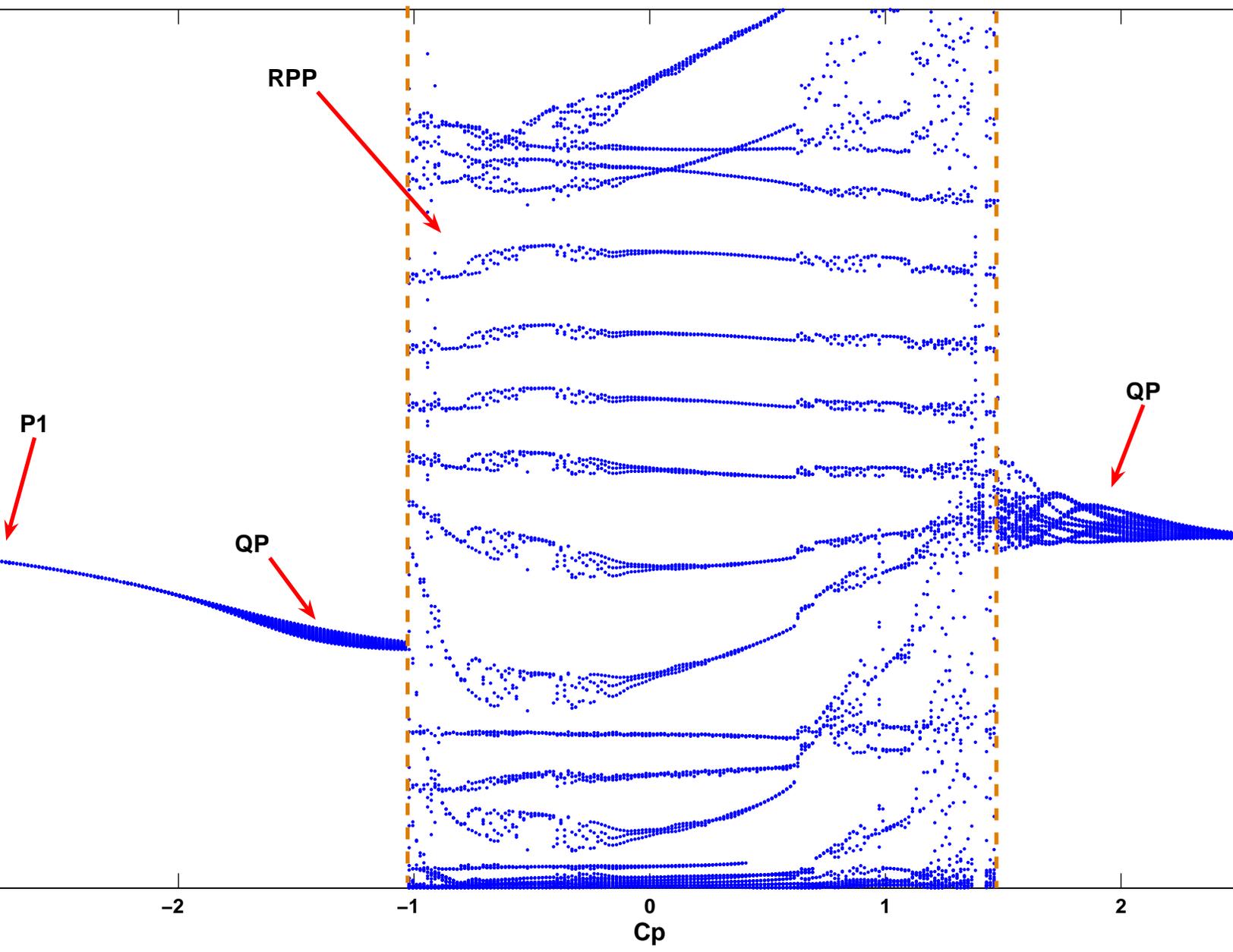
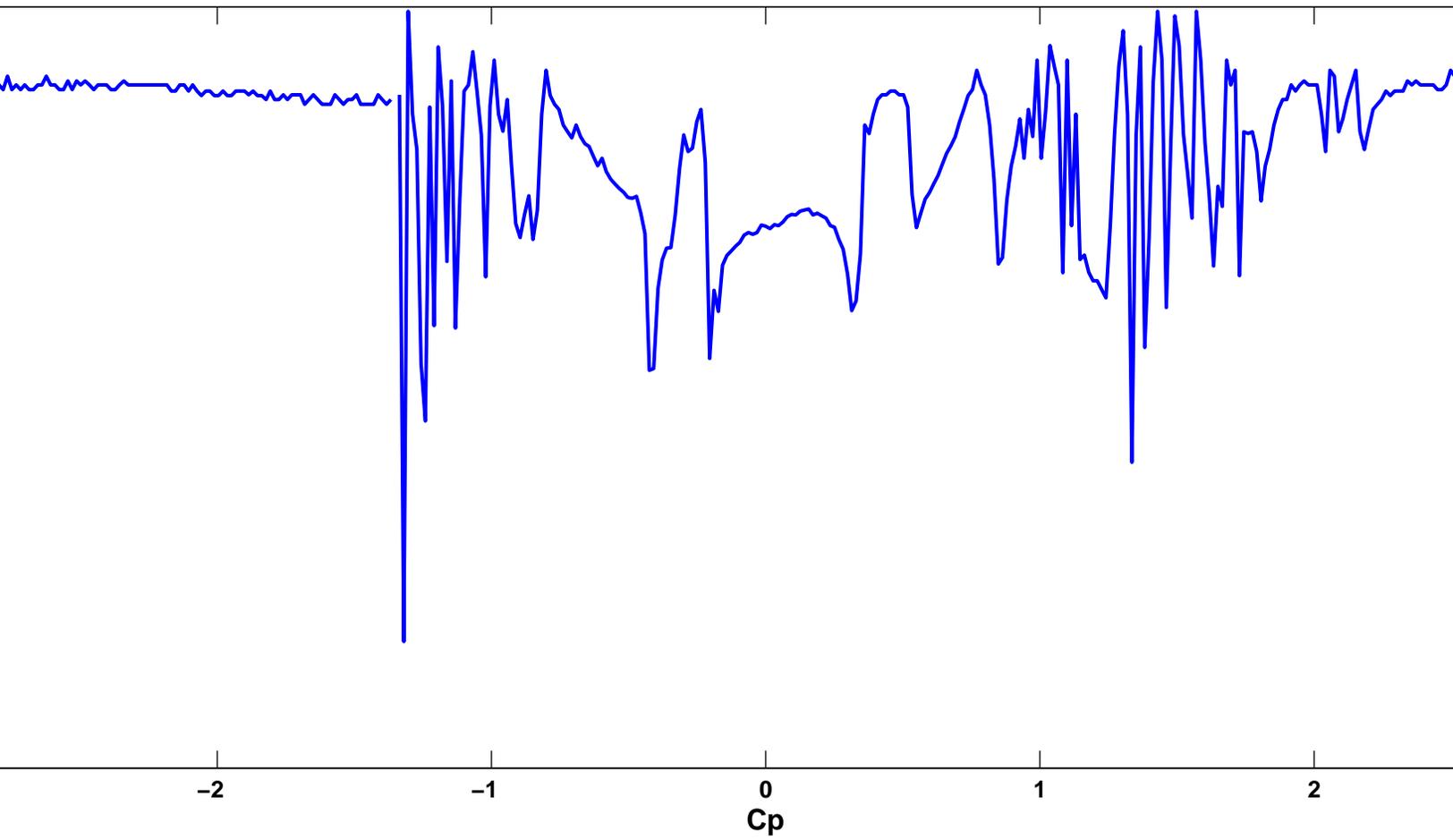
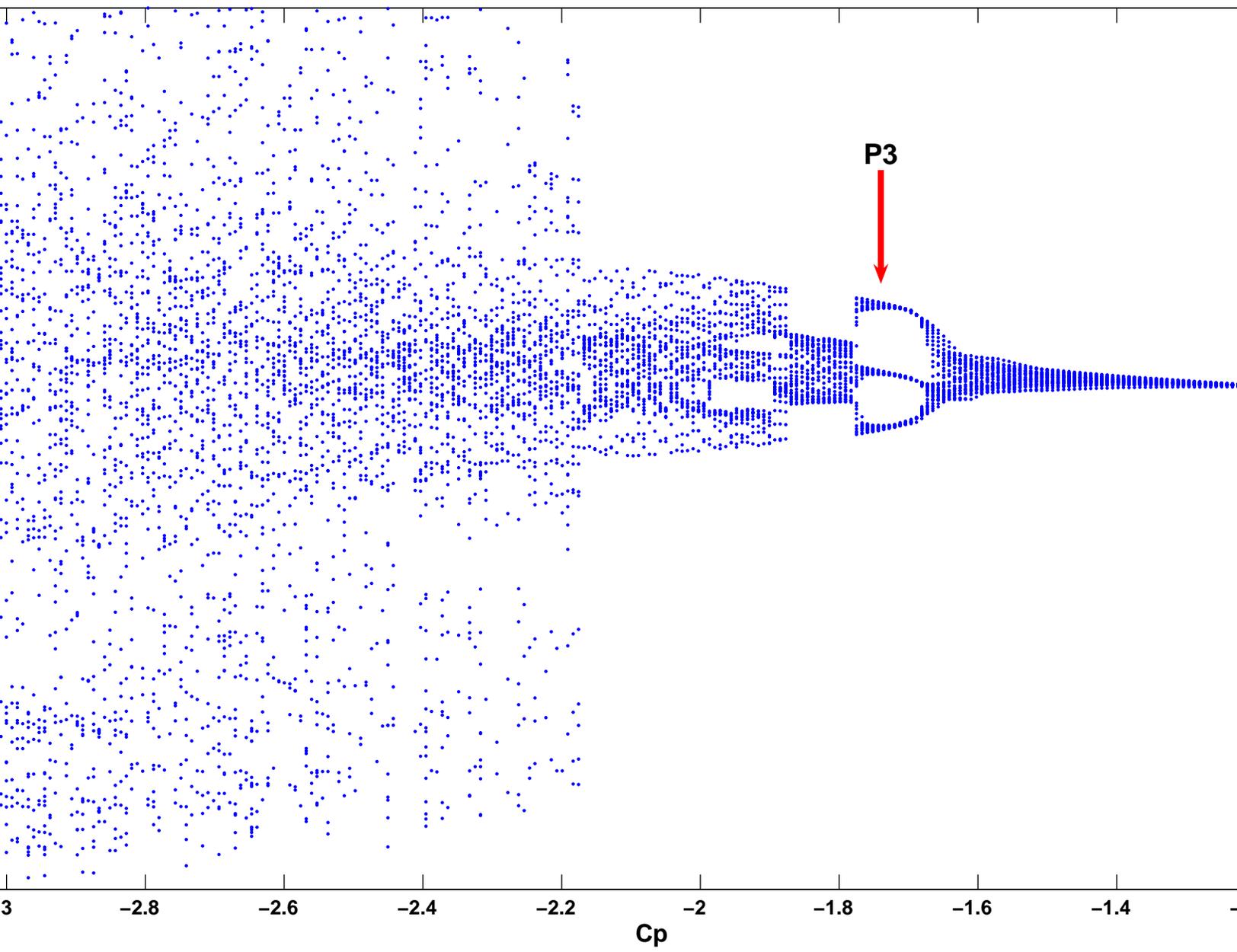
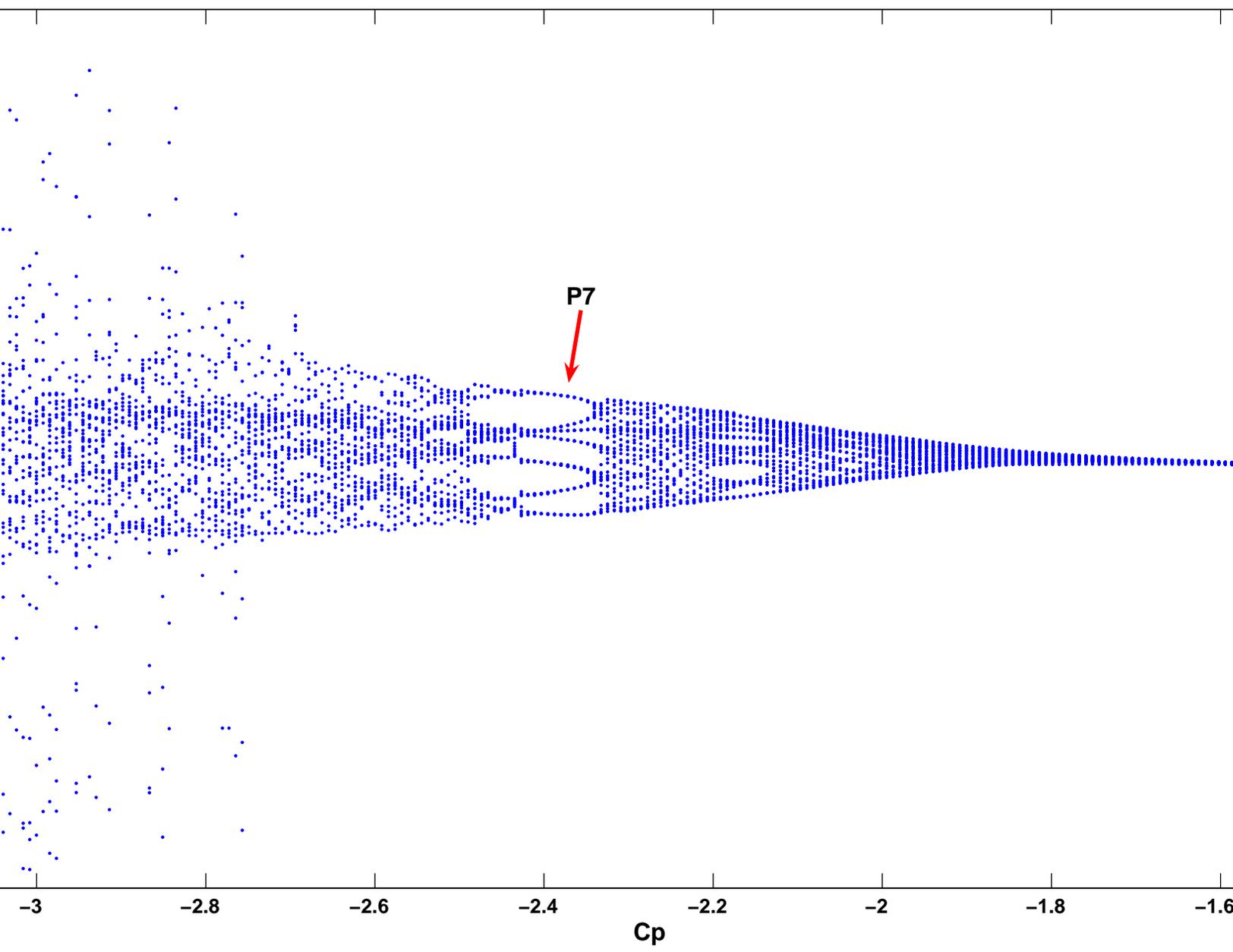
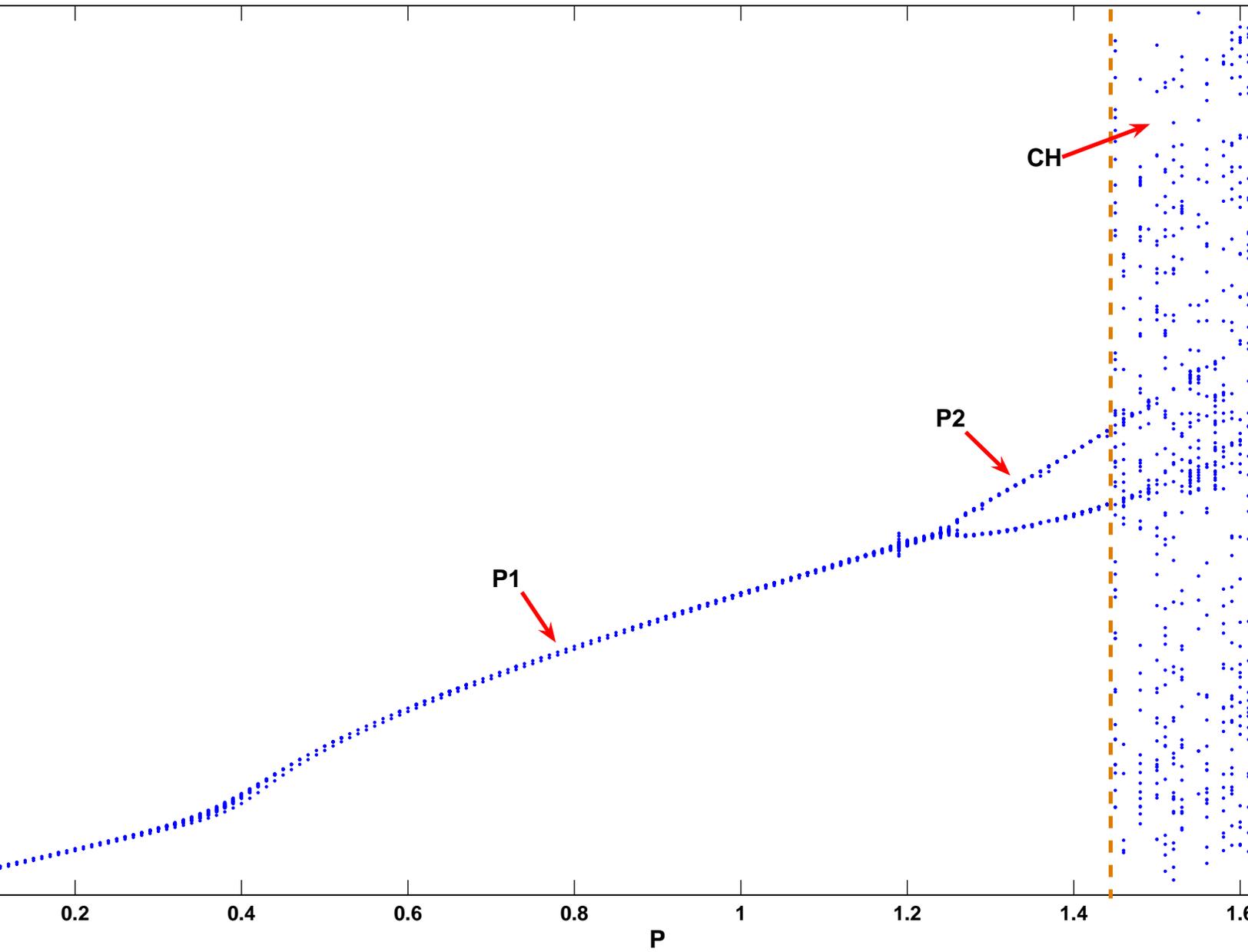
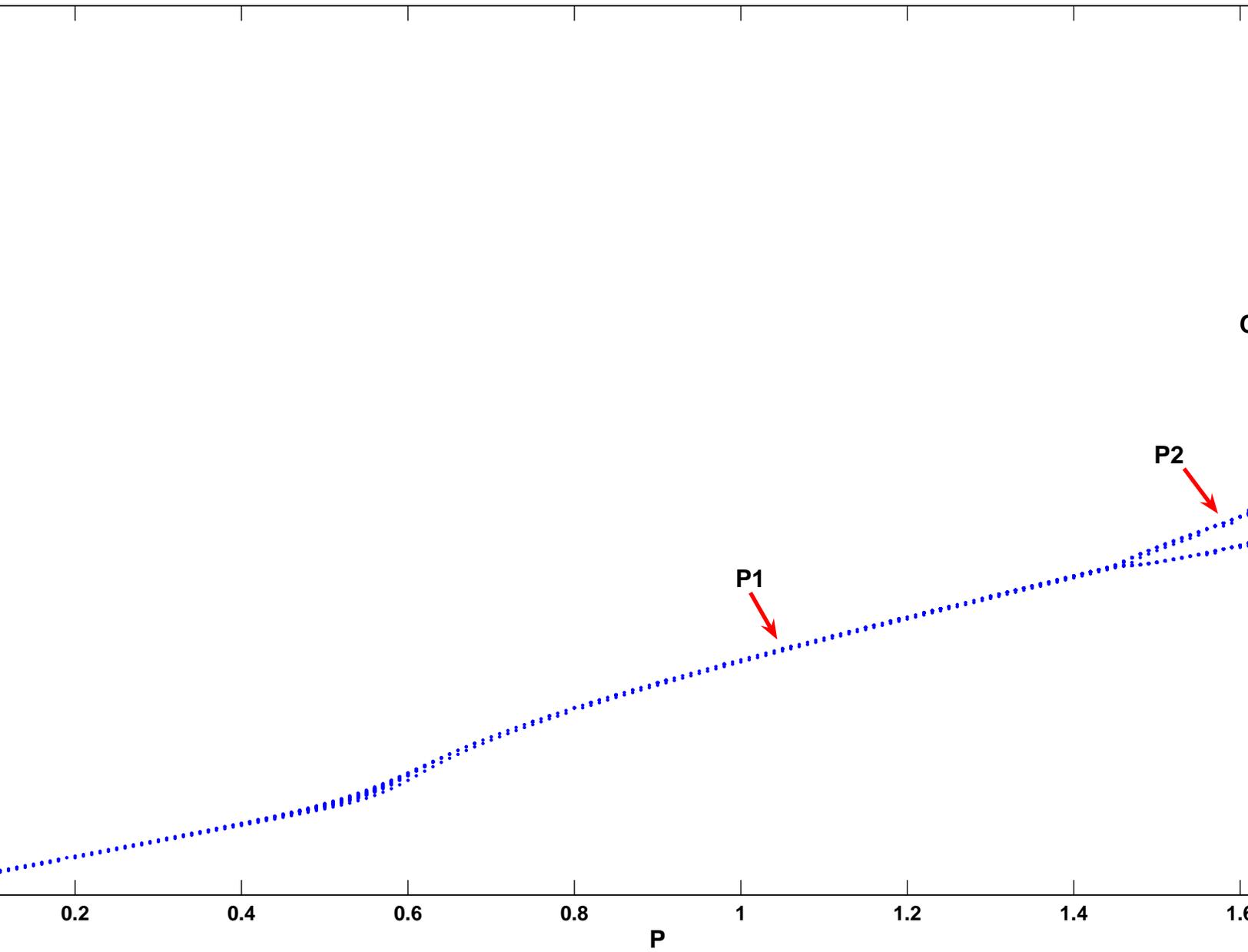
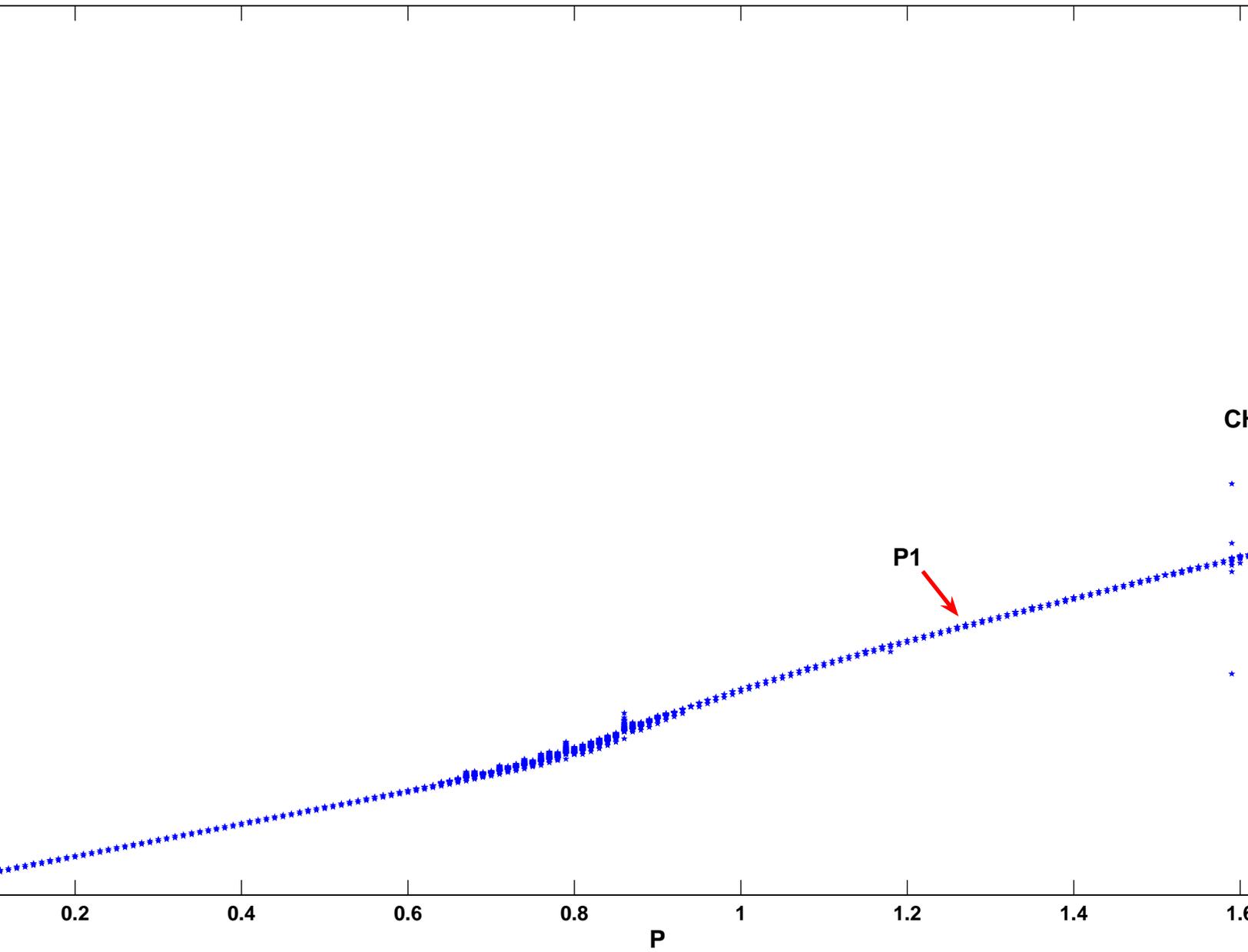
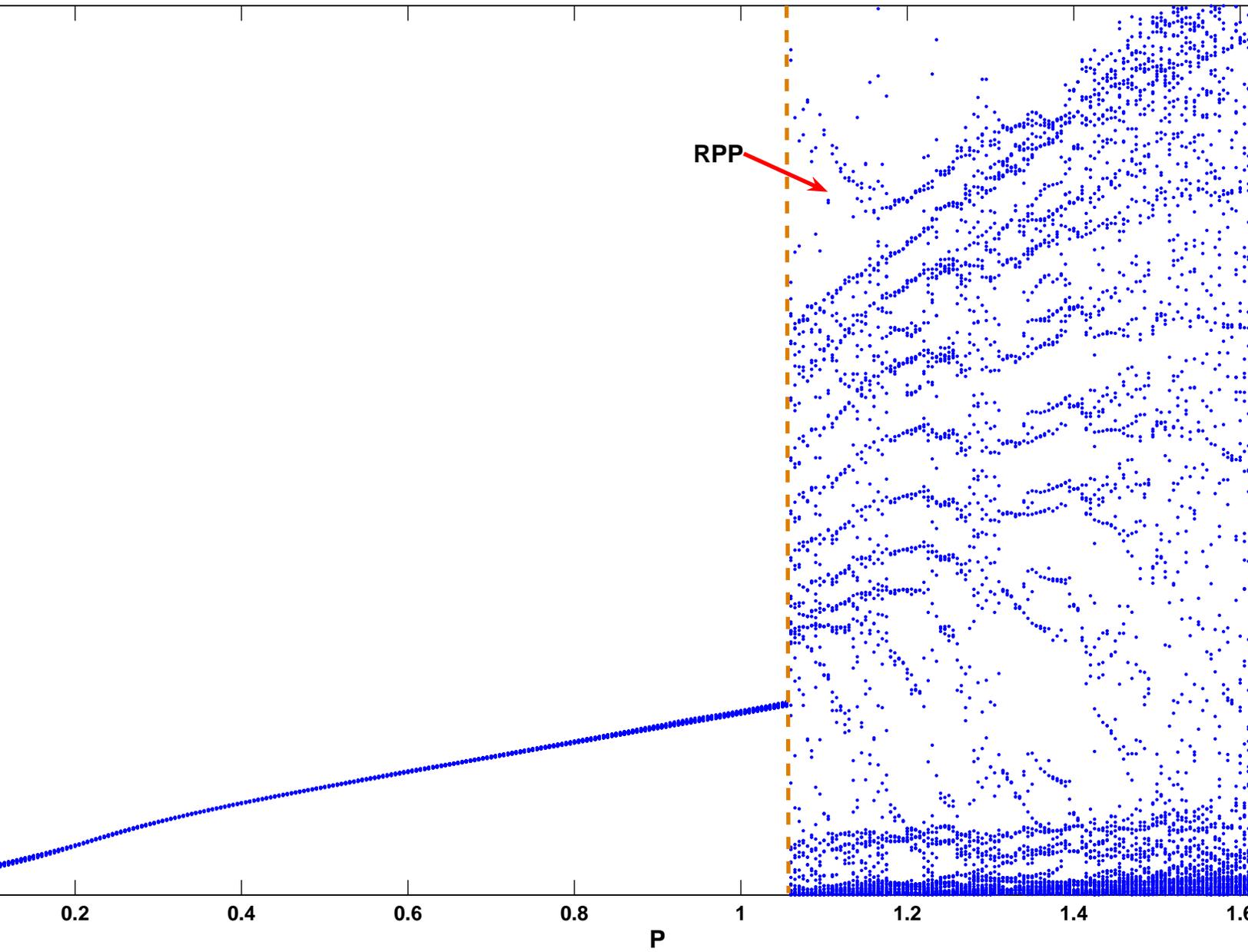
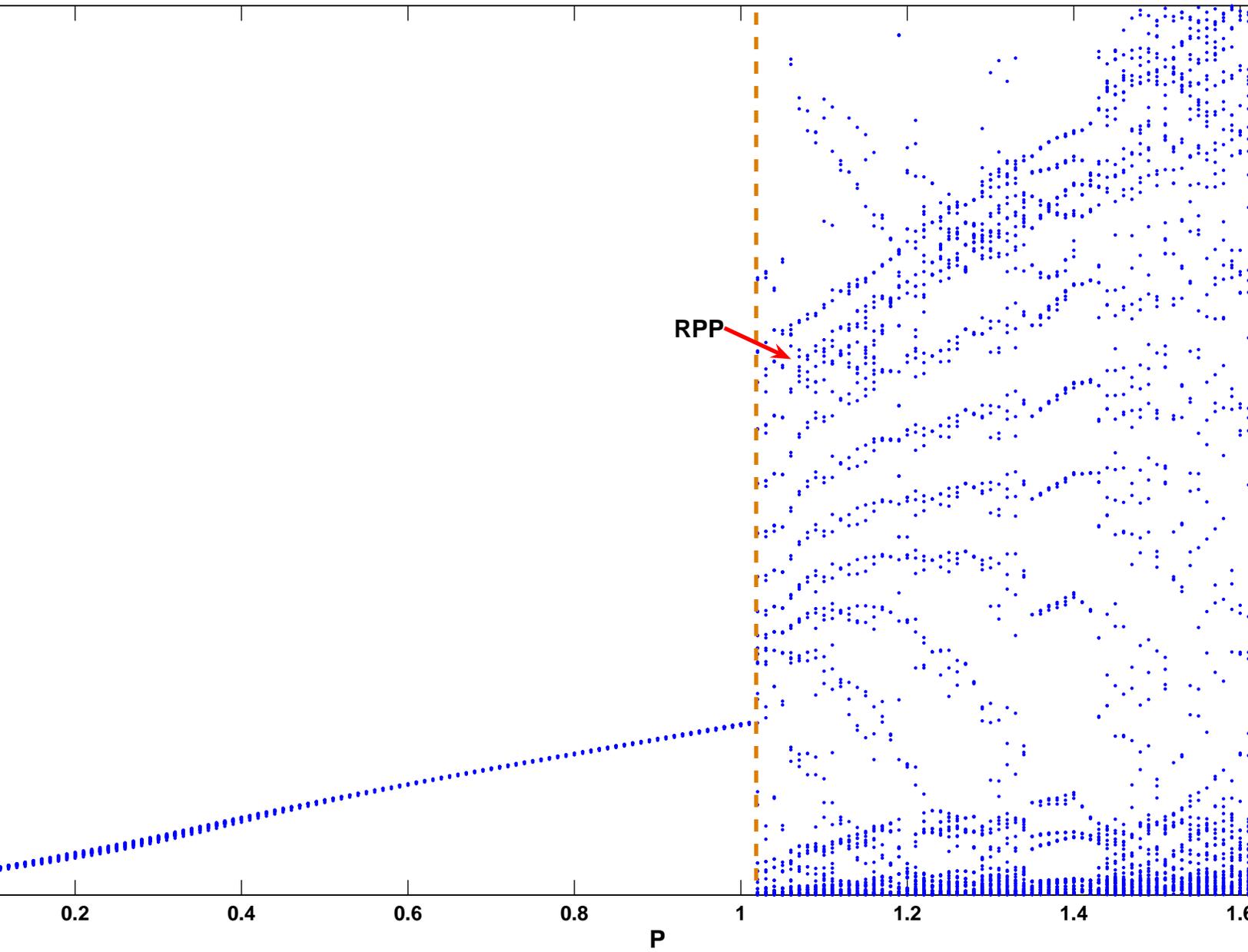
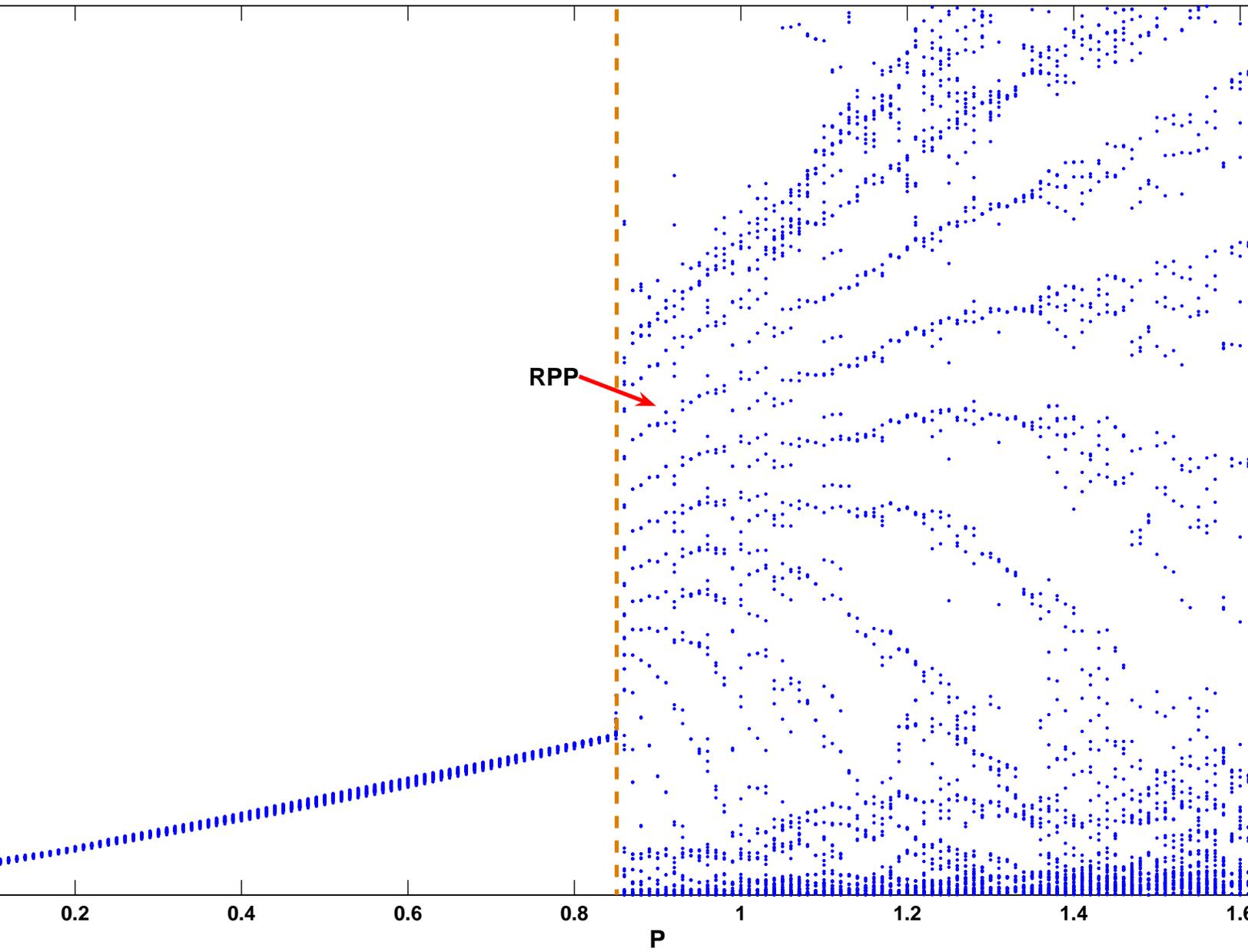
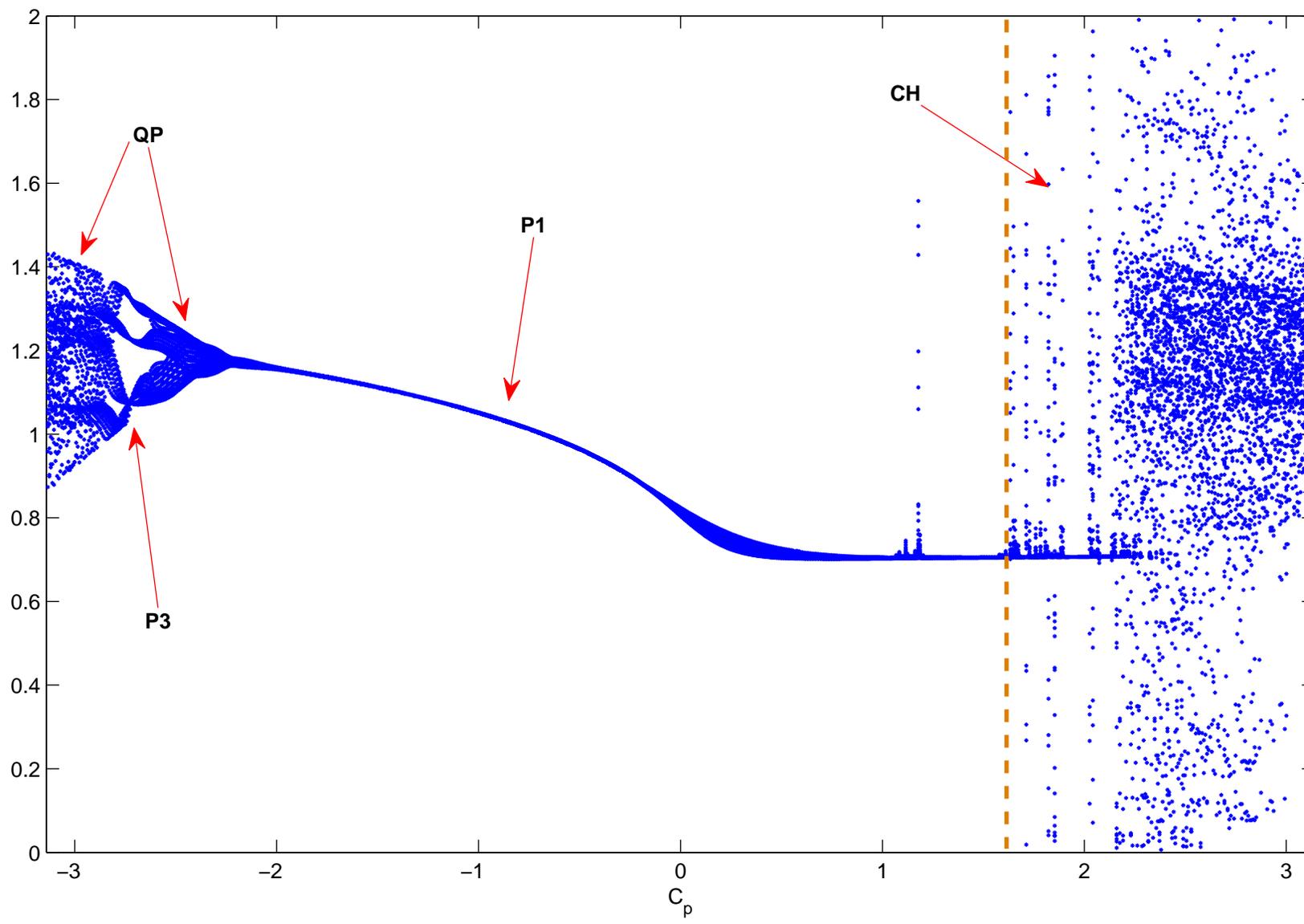
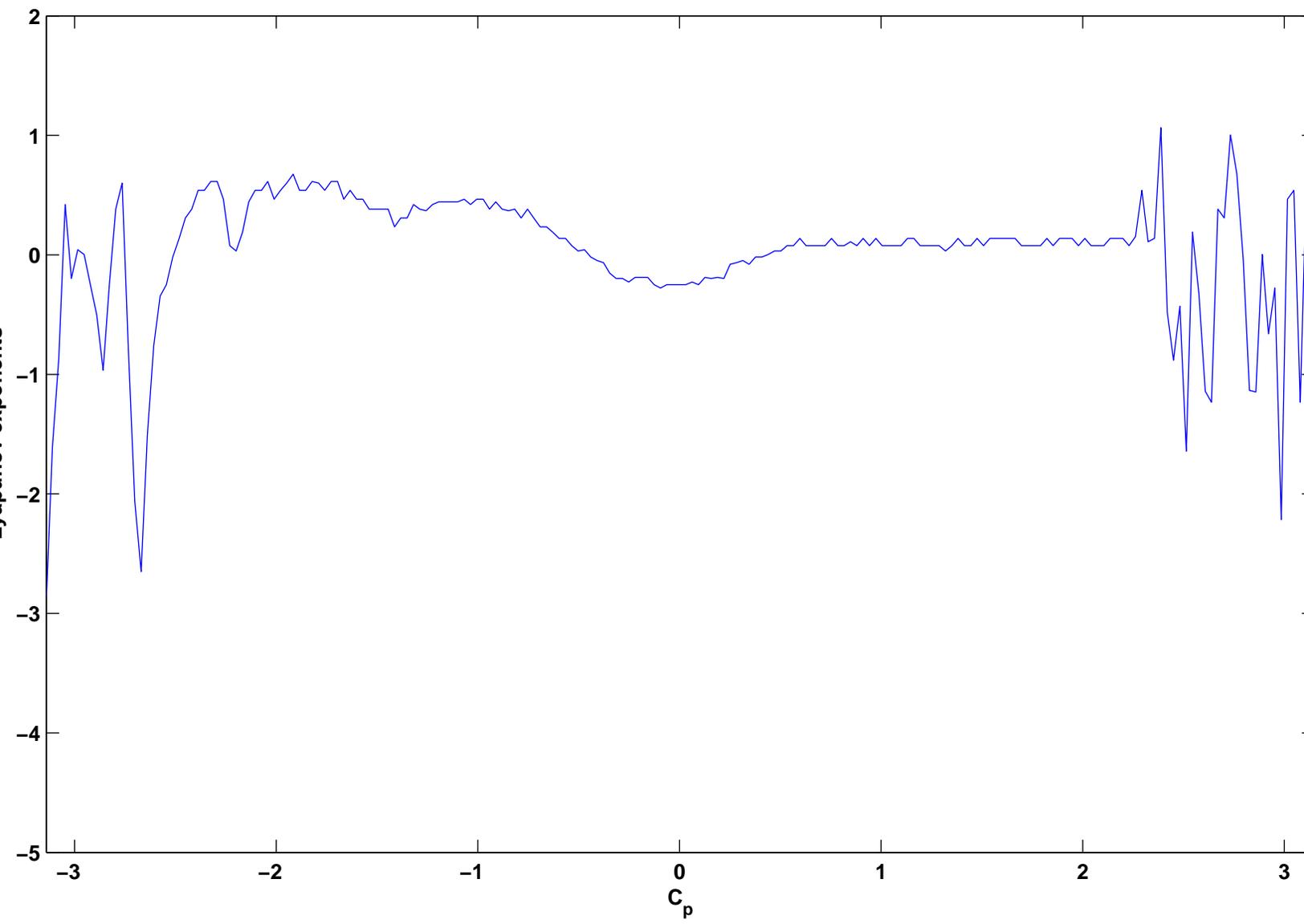
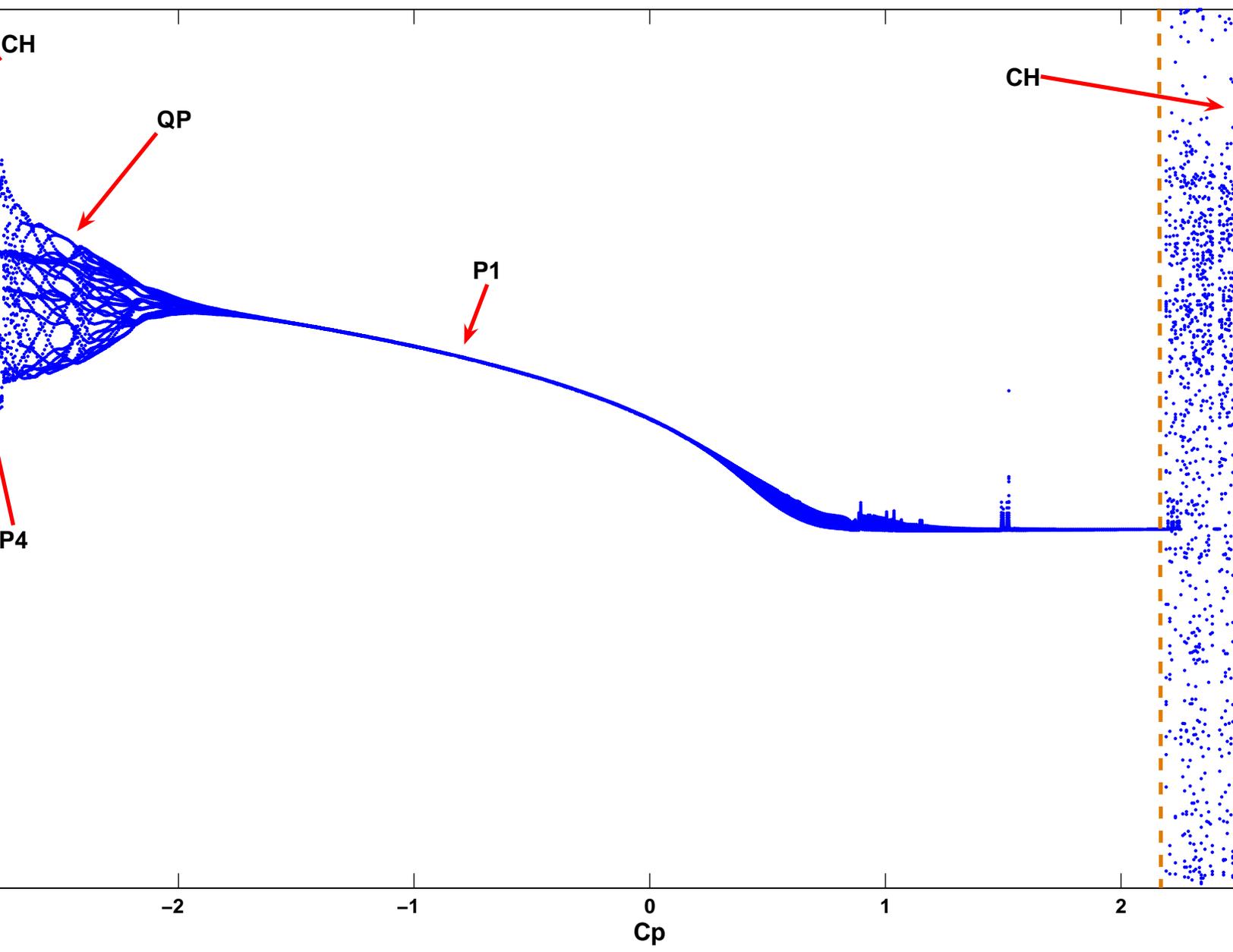
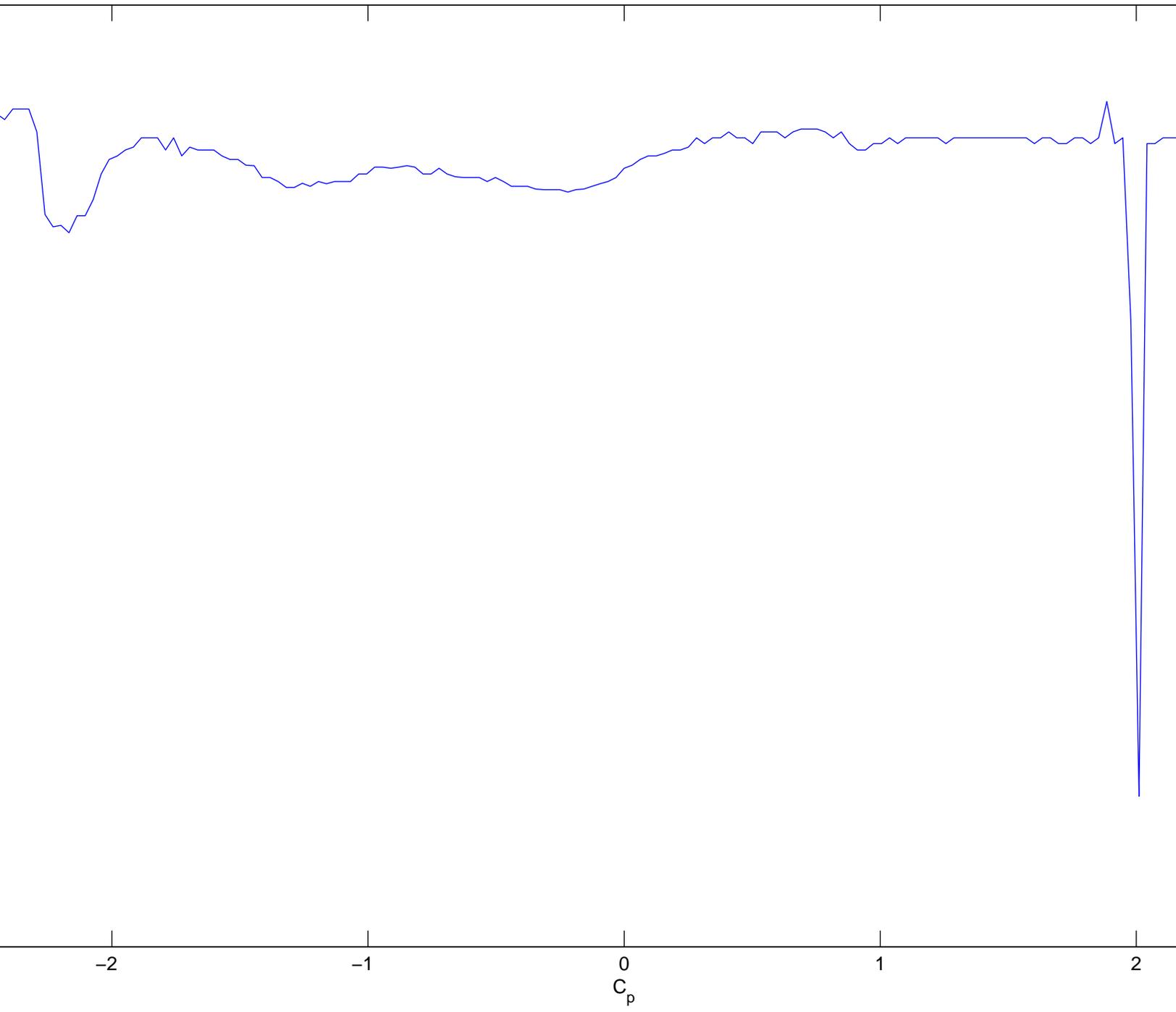
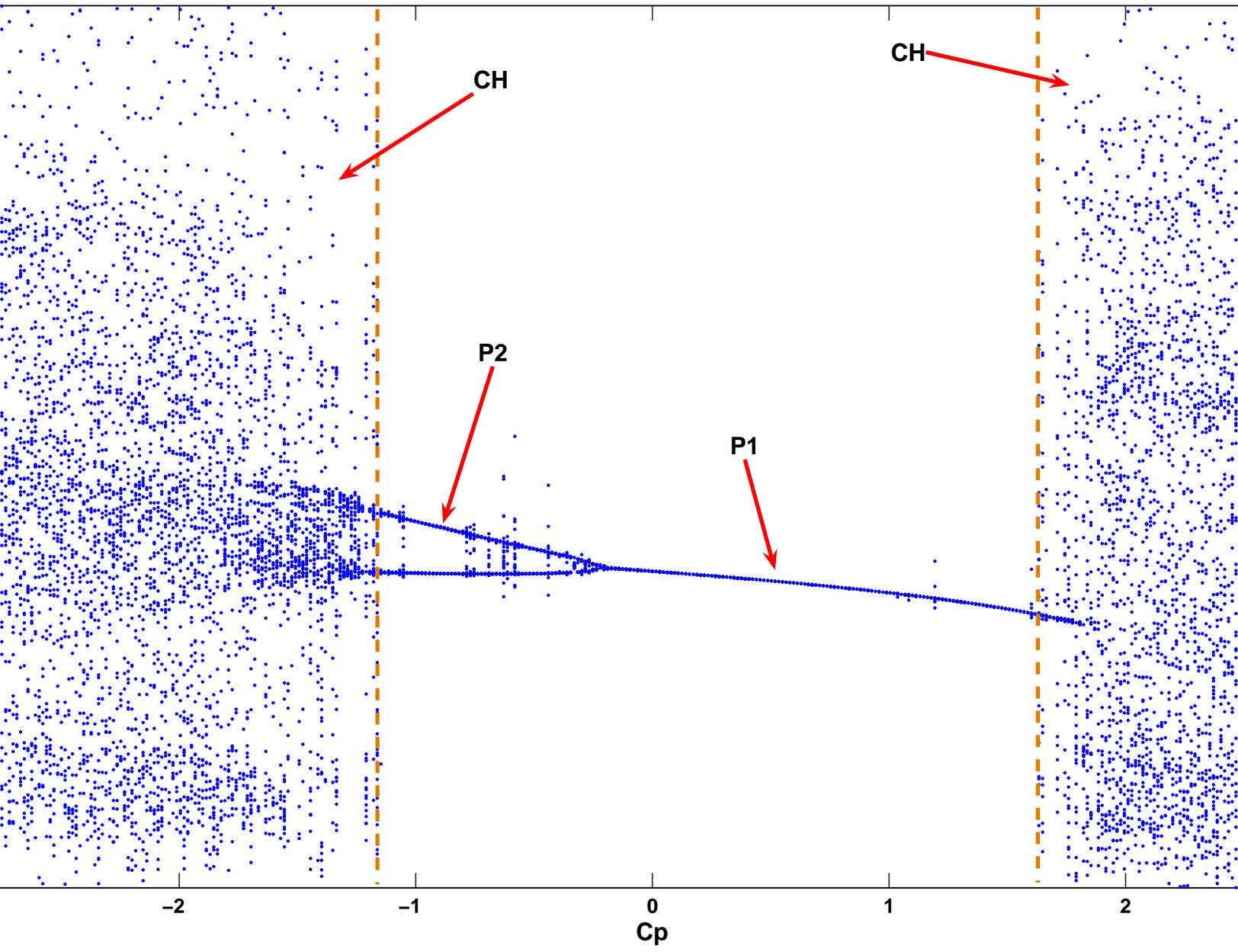
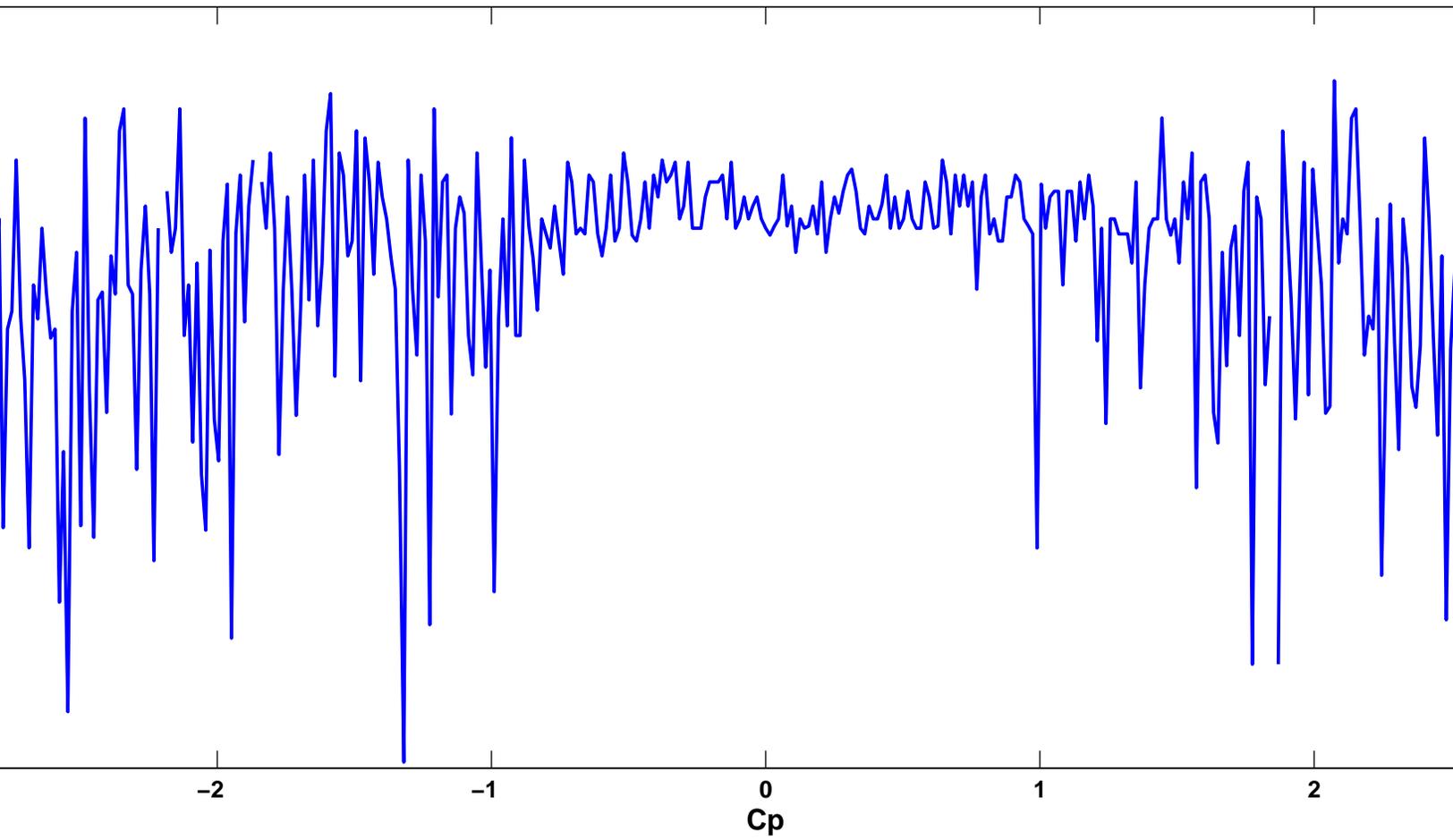
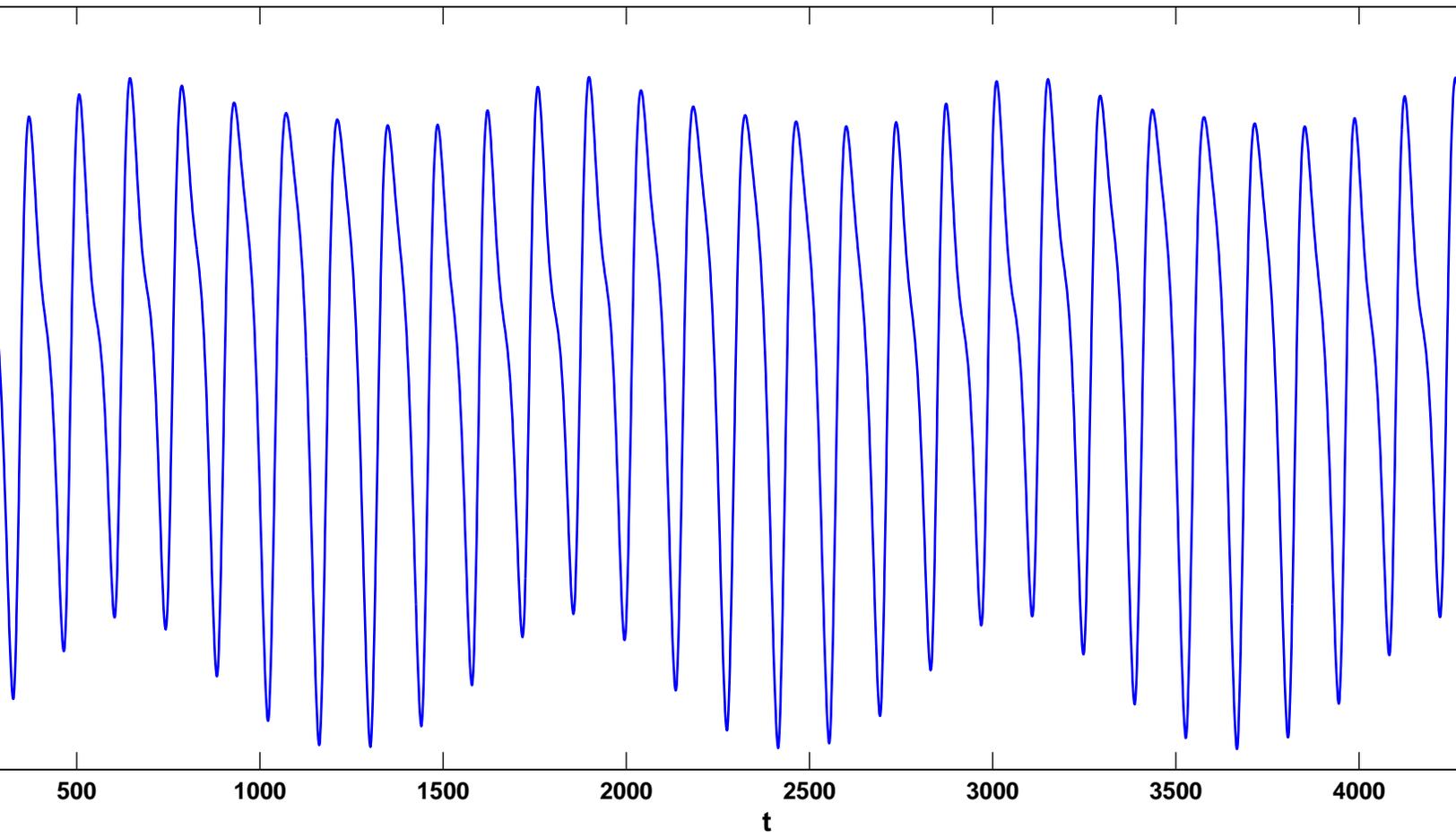
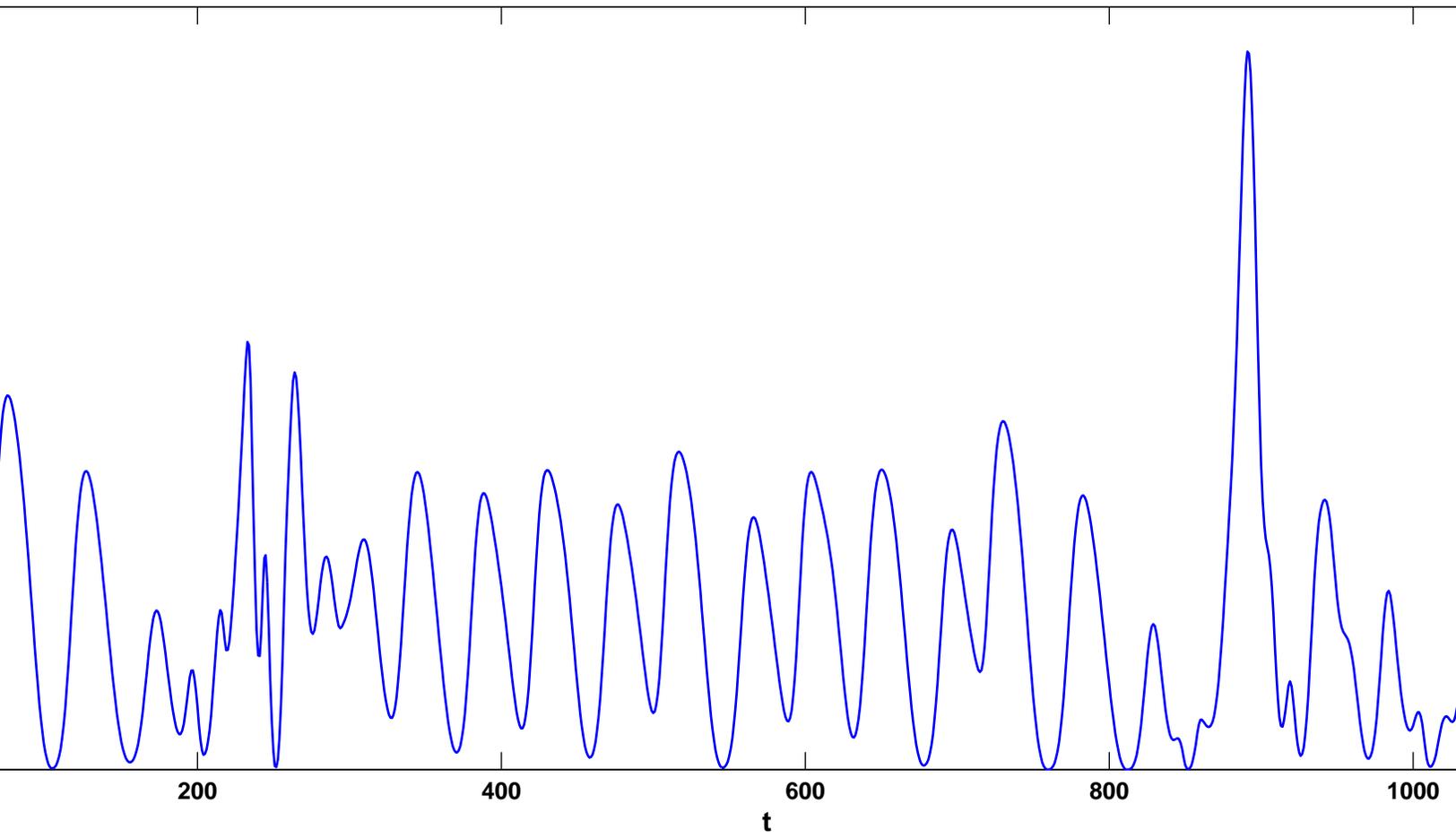
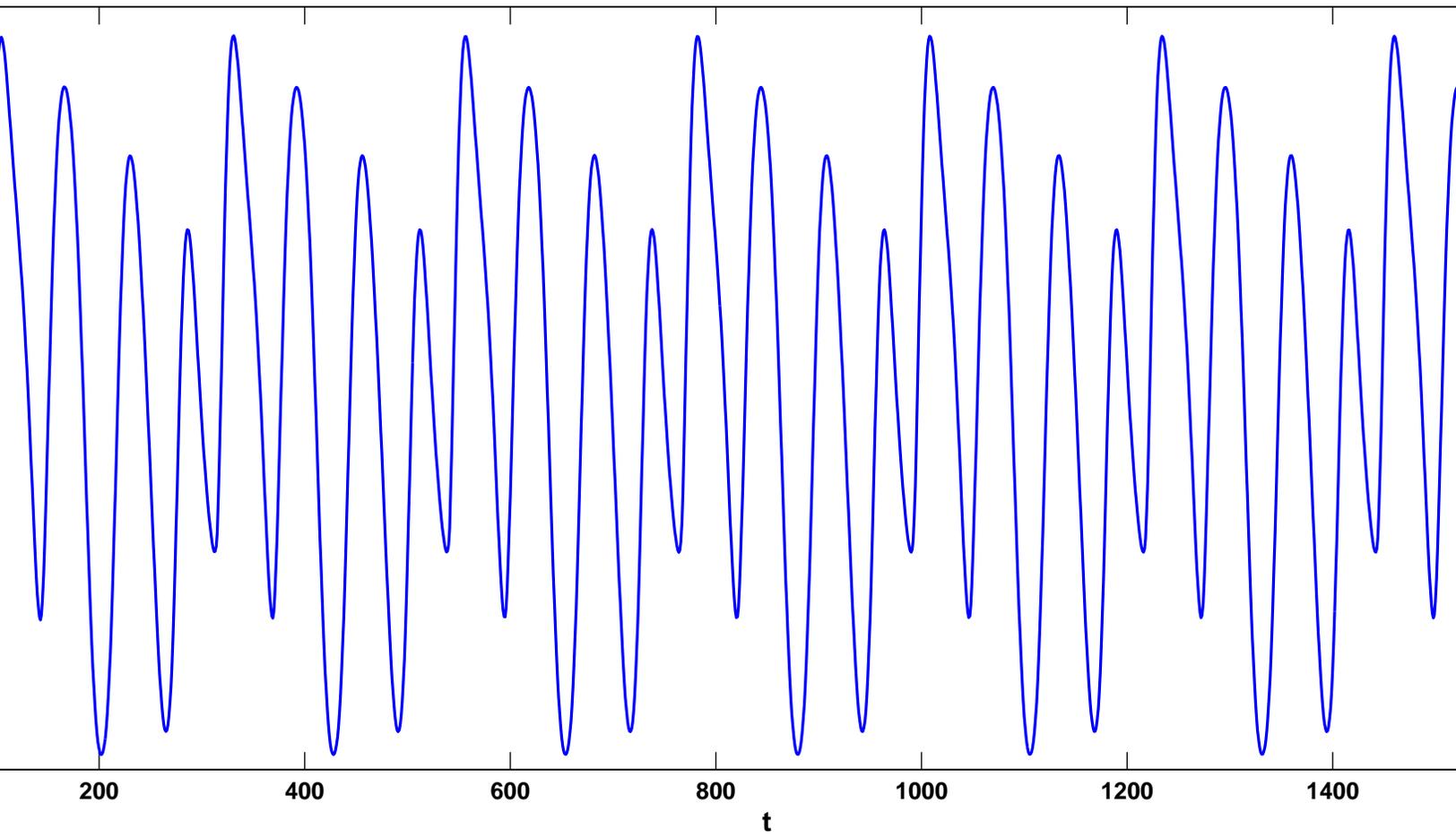
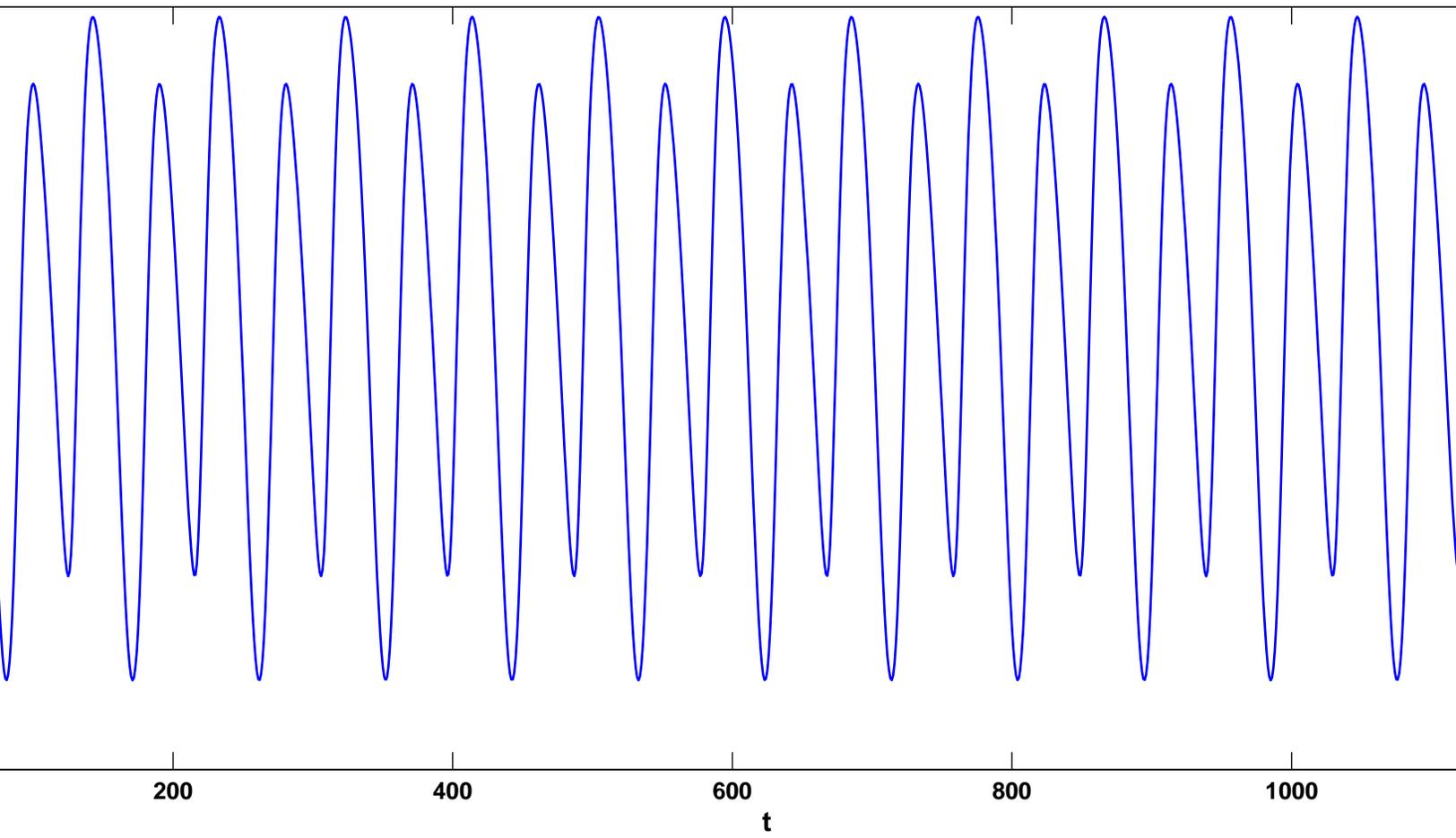
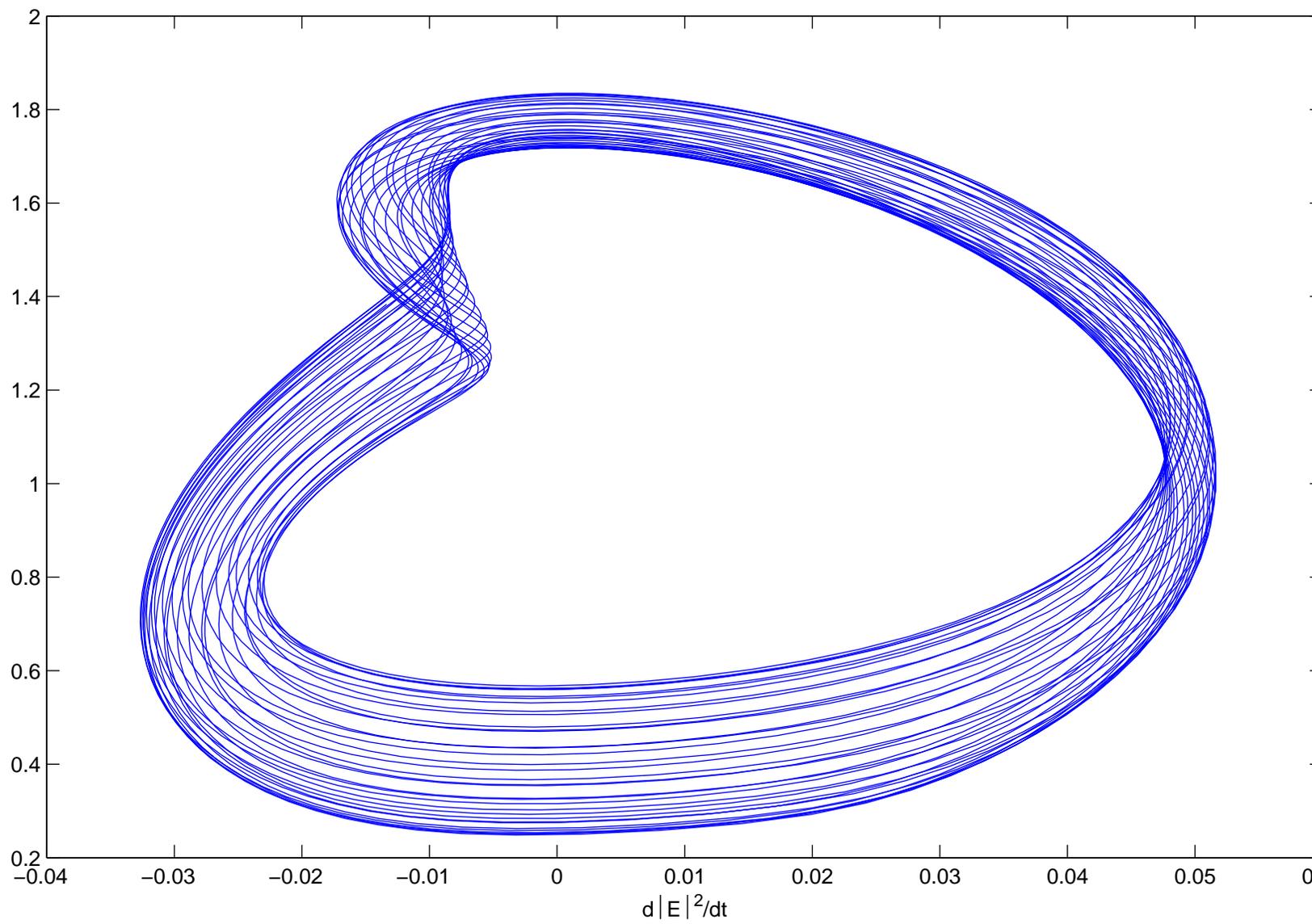
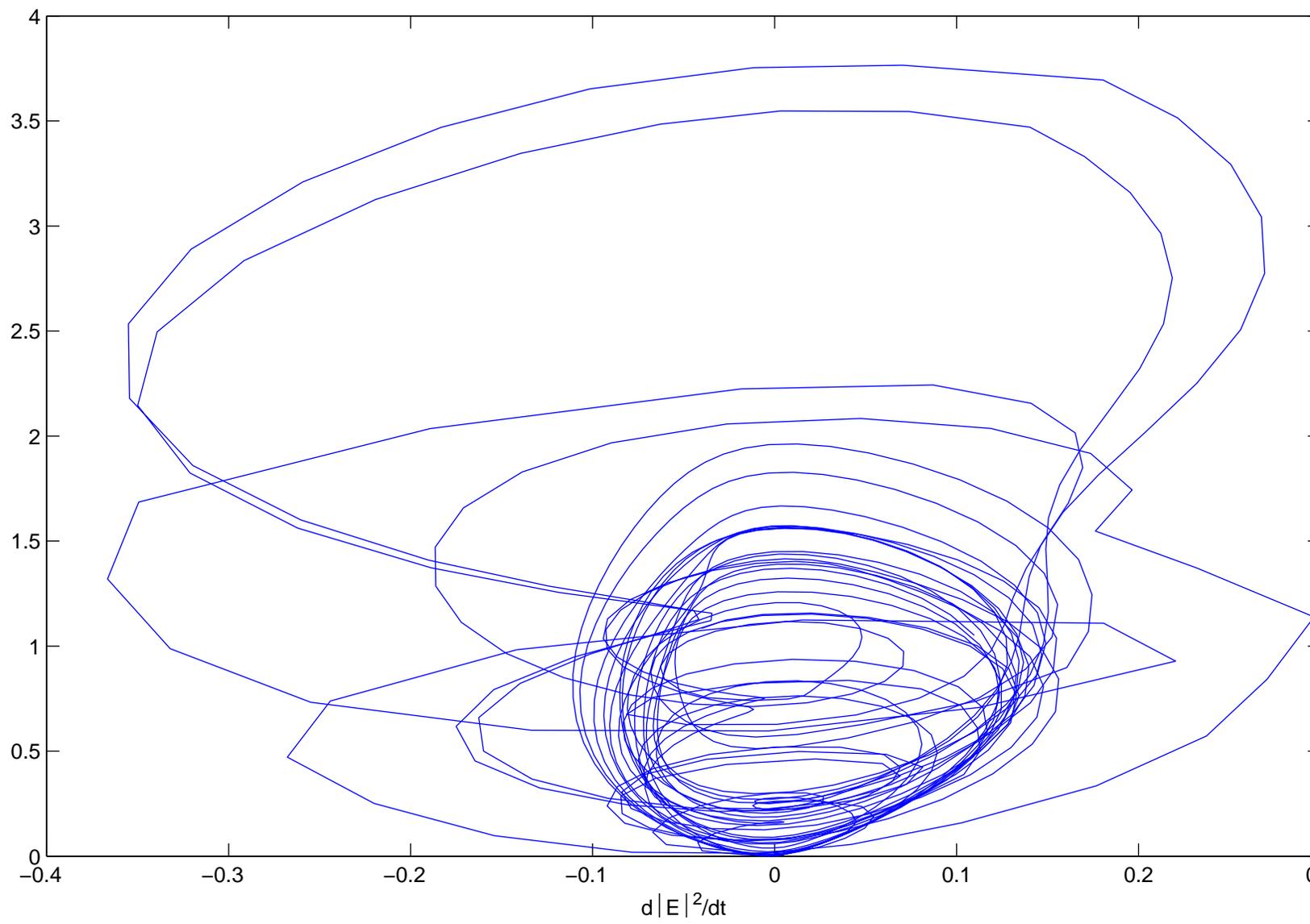
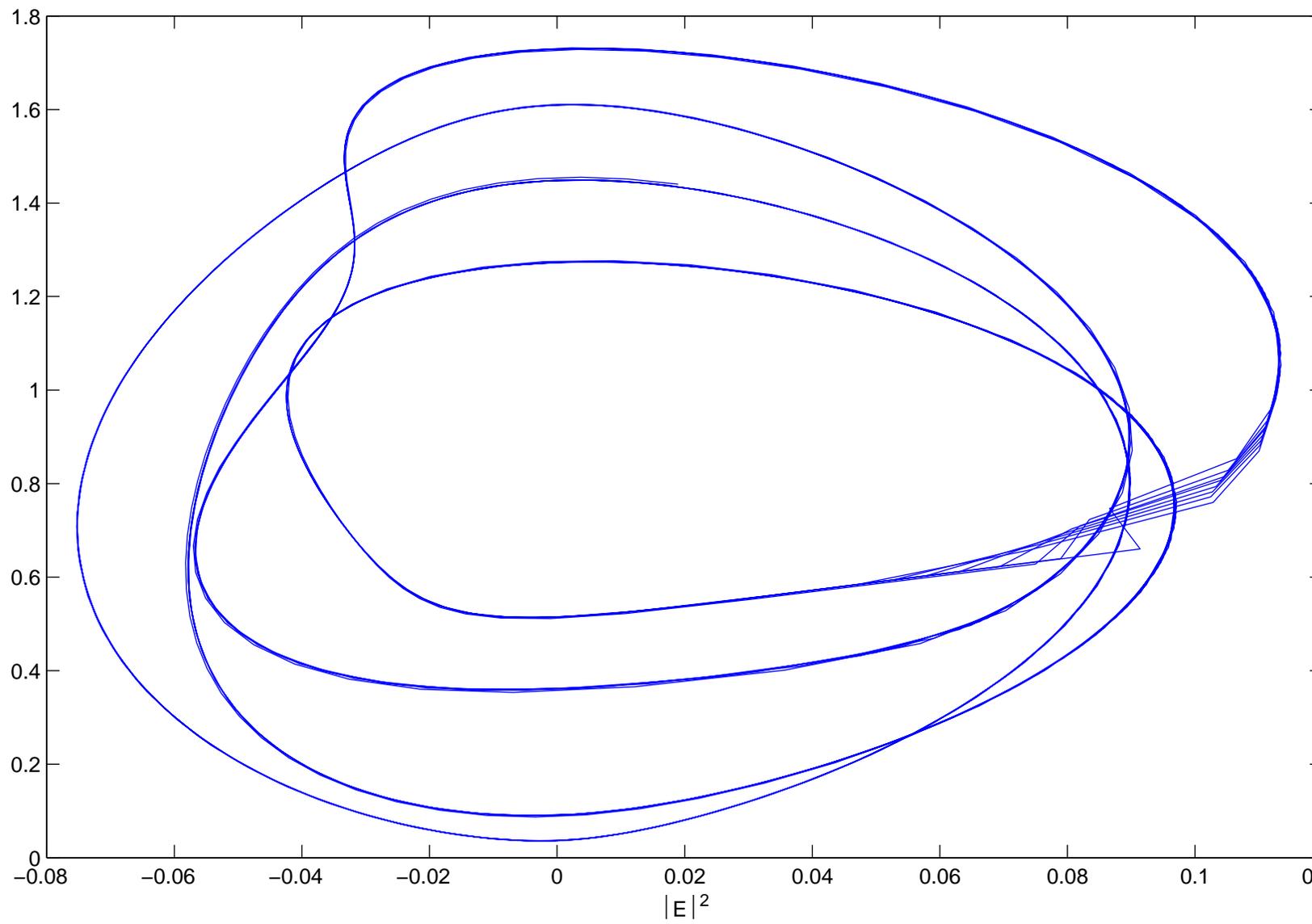
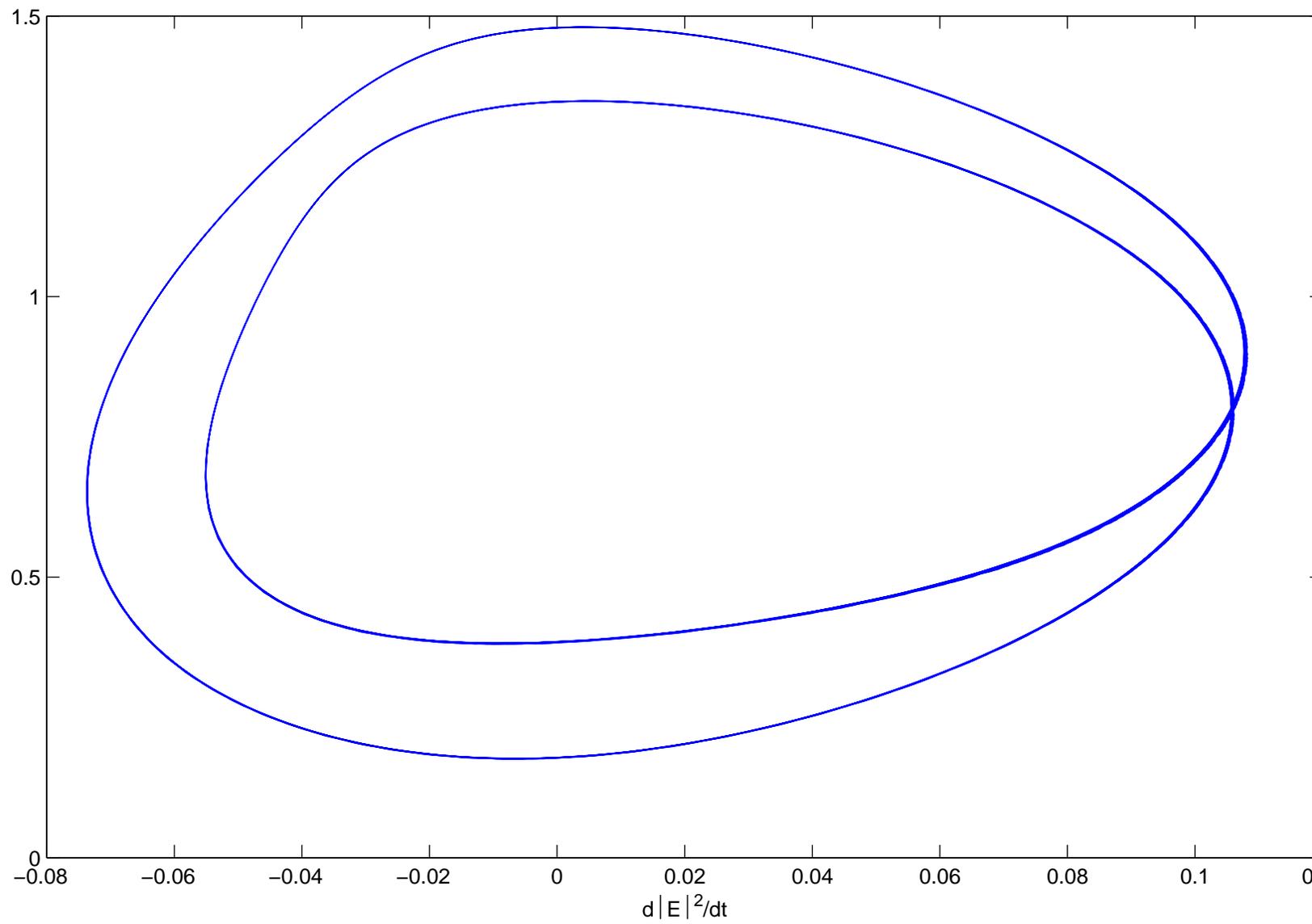
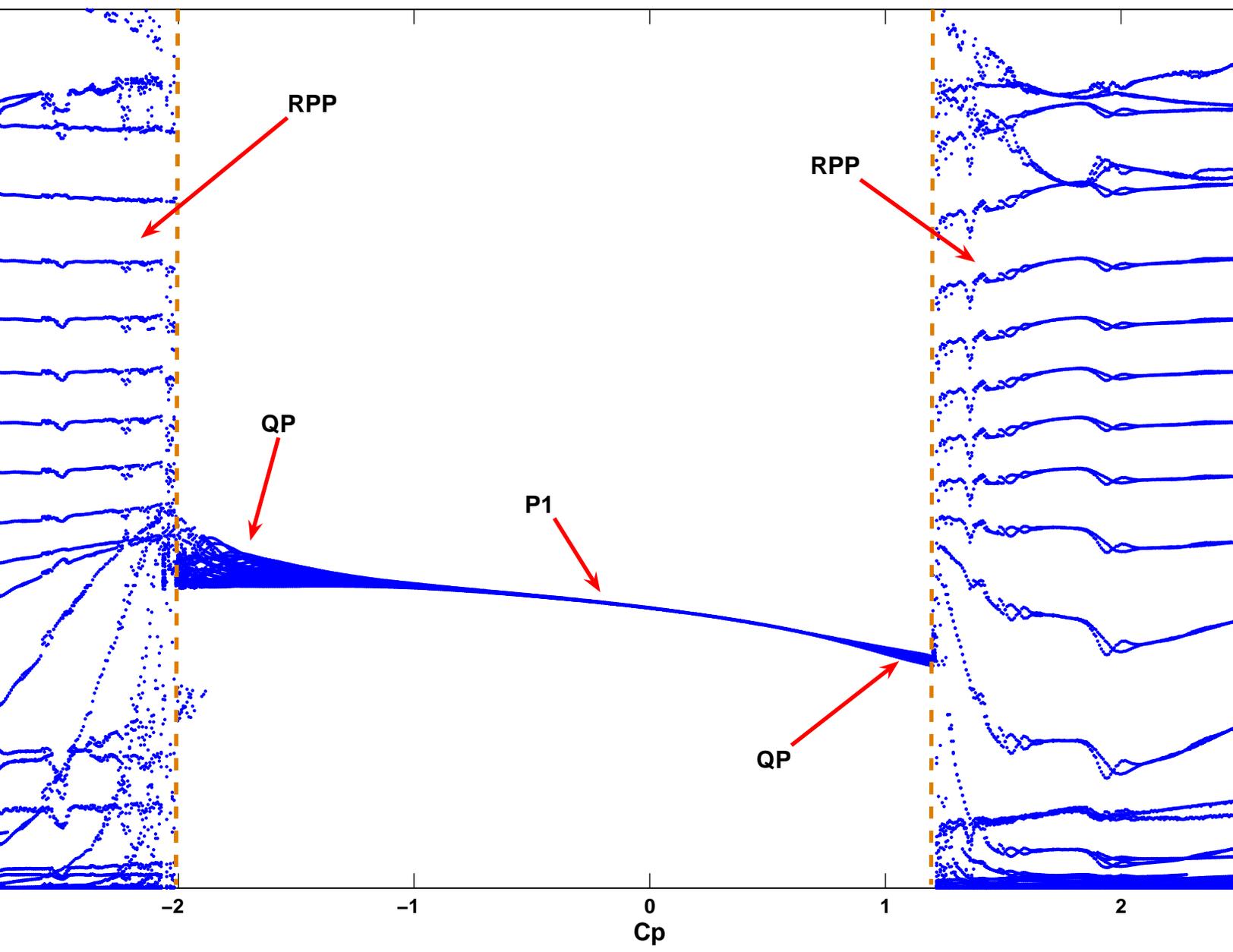
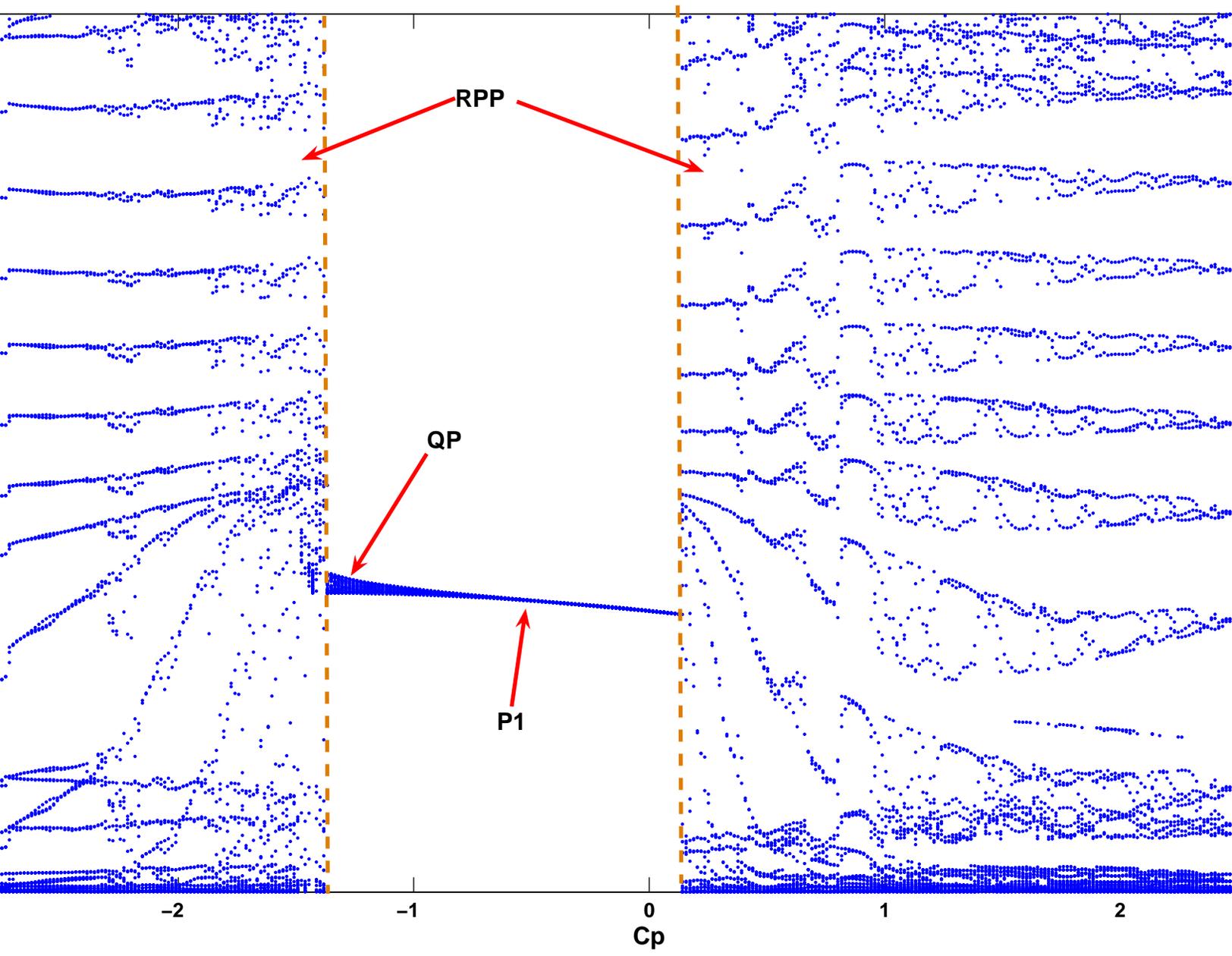
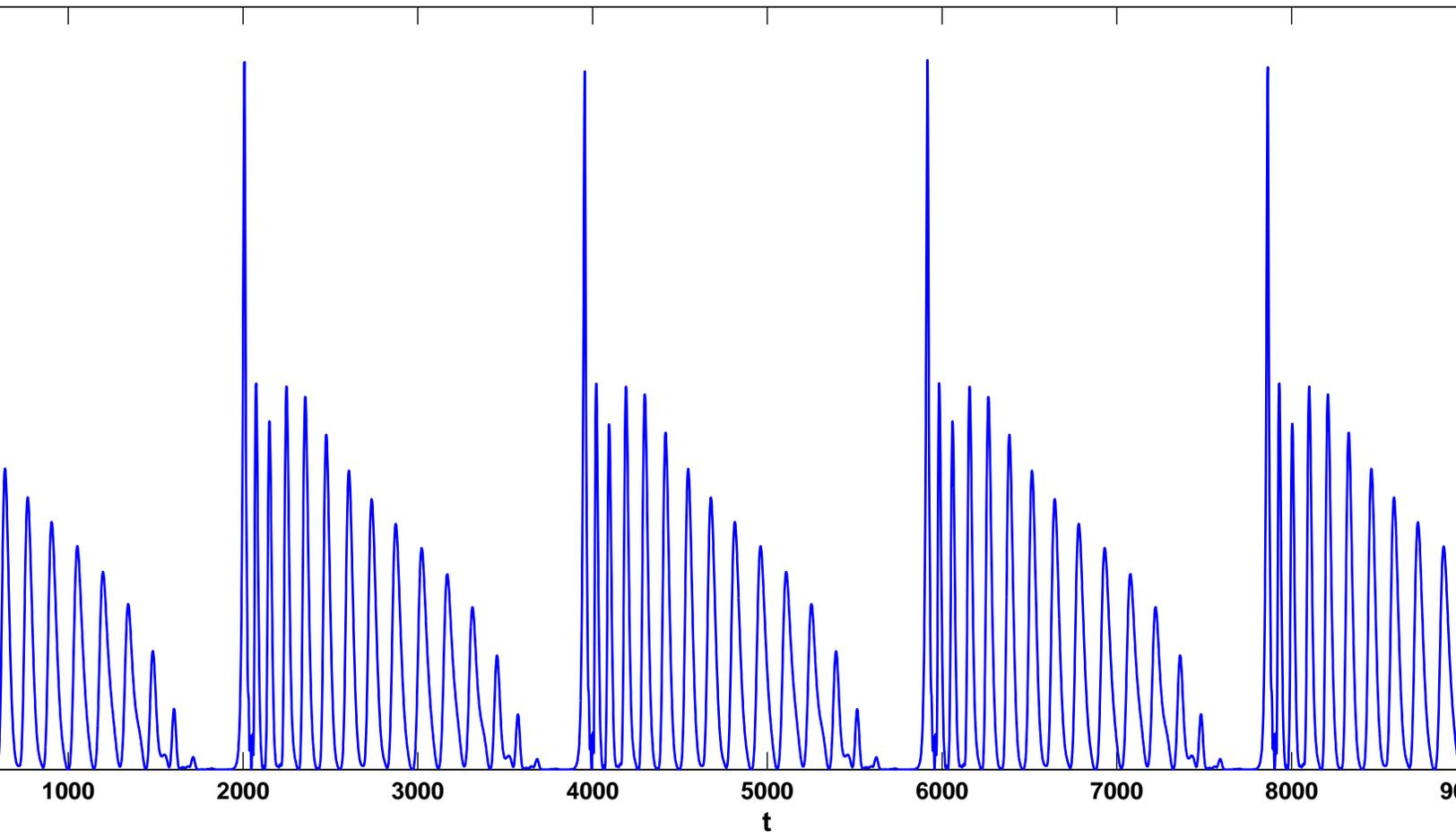
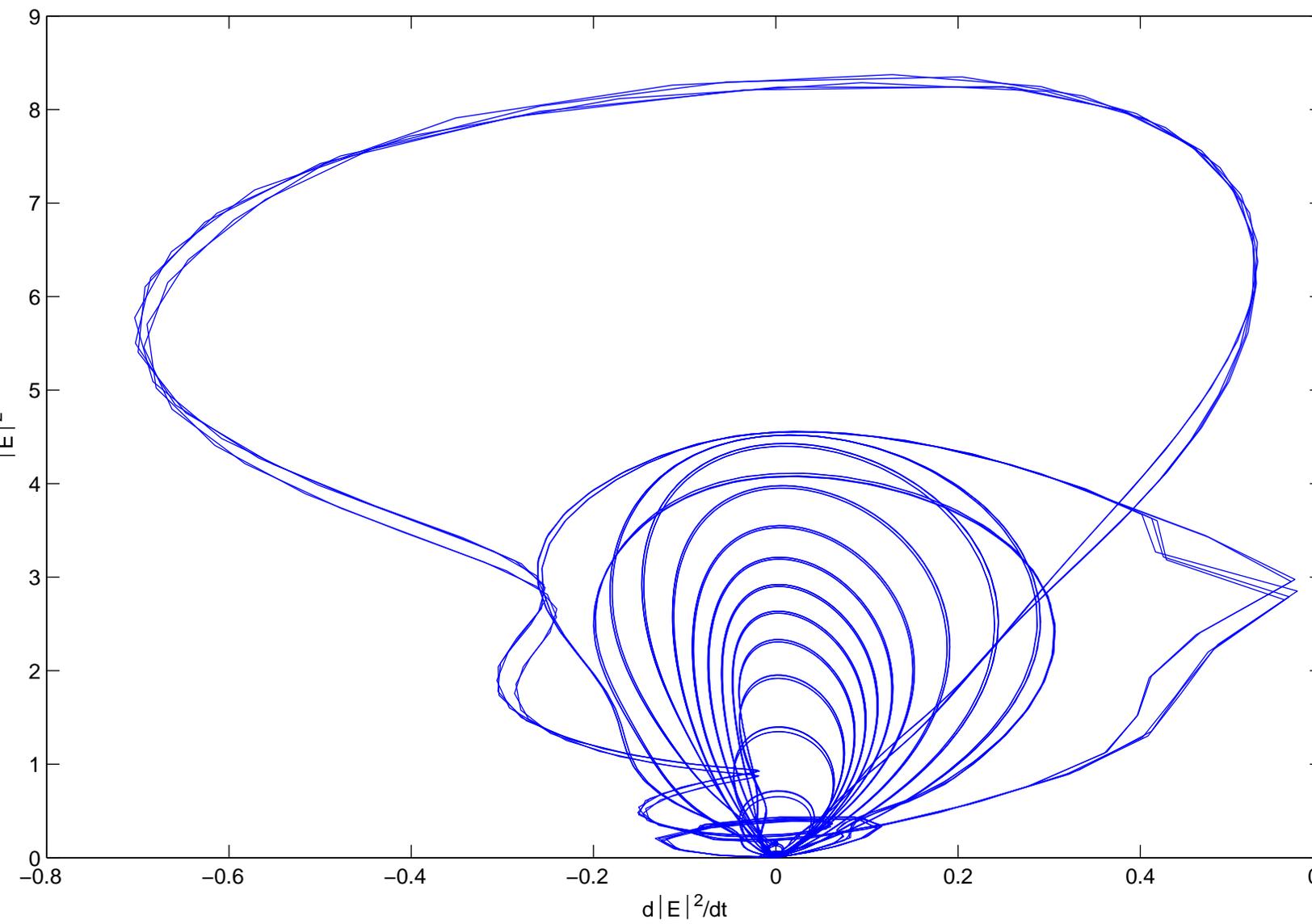
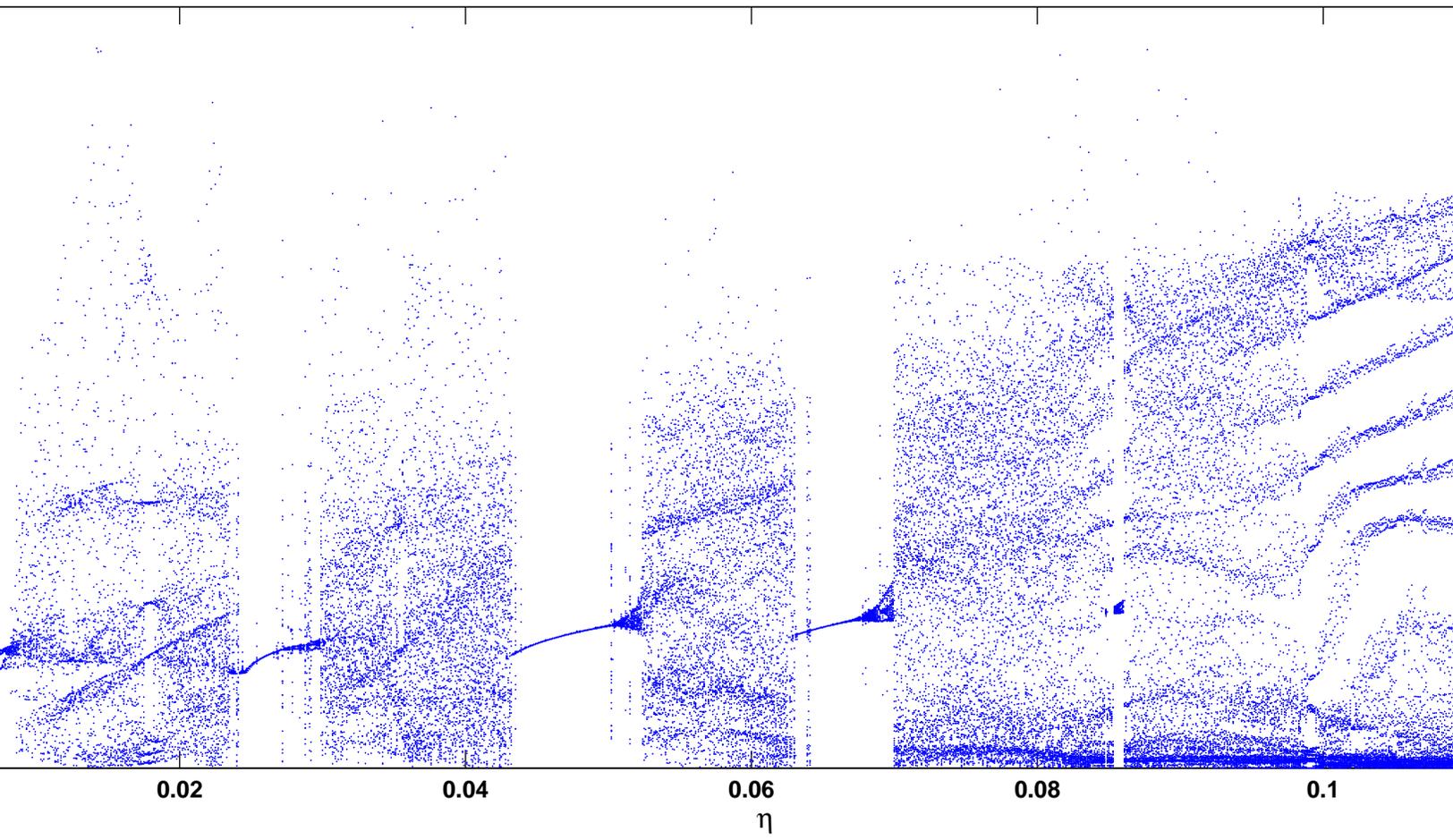
\end{document}